\documentclass[a4paper,fleqn,usenatbib,useAMS]{mnras}
\usepackage{graphicx}	
\usepackage{amsmath}	
\usepackage{amssymb}	
\usepackage{multicol}        
\usepackage{bm}		
\usepackage{pdflscape}	
\usepackage[T1]{fontenc}
\usepackage{ae,aecompl}

\newcommand{\lr}[1]{\left( #1 \right)}

\newcommand{\lrt}[1]{\left< #1 \right>}
\newcommand{\lrv}[1]{\left| #1 \right|}

\newcommand{\bbe}{{\boldsymbol \beta}}
\newcommand{\tbbe}{{\tilde{\bbe}}}
\newcommand{\bth}{{\boldsymbol \theta}}

\newcommand{\bga}{{\boldsymbol \gamma}}

\newcommand{\mbF}{{\mbox{\boldmath$F$}}}
\newcommand{\mbG}{{\mbox{\boldmath$G$}}}

\newcommand{\mbcF}{{\mbox{\boldmath$\cF$}}}
\newcommand{\mbcG}{{\mbox{\boldmath$\cG$}}}
\newcommand{\mbcH}{{\mbox{\boldmath$\cH$}}}
\newcommand{\mbcFEB}{\mbcF_{E}}
\newcommand{\mbcGEB}{\mbcG_{E}}
\newcommand{\mbcFEC}{\mbcF_{E'}}
\newcommand{\mbcGEC}{\mbcG_{E'}}

\newcommand{\mbg}{{\mbox{\boldmath$g$}}}

\newcommand{\Real}[1]{{\rm Re}\left[ #1 \right]}
\newcommand{\PP}[2]{{\frac{\partial {#1}}{\partial {#2}}}}

\newcommand{\cM}{{\cal M}}

\newcommand{\cG}{{\cal G}}
\newcommand{\cH}{{\cal H}}

\newcommand{\cF}{{\cal F}}

\usepackage{newtxtext,newtxmath}

\title{New highly precise weak gravitational lensing flexions measurement method based on ERA method.}

\author[Yuki Okura, Toshifumi Futamase]
{Yuki Okura,$^{1,2}$\thanks{E-mail: yuki.okura@nao.ac.jp}
Toshifumi Futamase,$^{3}$
\\
$^{1}$NAOJ\\
$^{2}$RIKEN\\
$^{3}$Kyoto Sangyo University}
\begin{document}
\label{firstpage}
\pagerange{\pageref{firstpage}--\pageref{lastpage}}
\maketitle

\begin{abstract}
Weak gravitational lensing flexions are a kind of weak lensing distortion which are defined as the spin 1 and spin 3 combinations of the third order derivatives of gravitational lensing potential. Since the shear has spin 2 combination of the second order derivative,  the flexion signal gives partly independent information from shear signal and is more sensitive to the local mass distribution than shear signal.
Thus its measurement is expected to play important roles in observational cosmology.    
However, since the weakness of the flexion signal as well as the complicatedness of its intrinsic noise made its accurate observation very difficult. We propose a new method of measuring the flexion signal using ERA method which is a method to measure weak
lensing shear without any approximation.
We find two particular combinations of the flexions which provide the quantities with only lensing information and free of intrinsic noise when taken average.
It is confirmed by simple numerical simulation that the statistical average of these combinations do not in fact depend on the strength of the intrinsic distortion.  
\end{abstract}
\onecolumn
\section{Introduction}
It is widely recognized that weak gravitational lensing shear analysis is a unique and powerful tool to analyze the mass distribution of the universe. Coherent deformation of the shapes of background galaxies carries not only the information of intervening mass distribution but also the cosmological background geometry and thus the cosmological parameters (Mellier 1999; Schneider 2006; Munshi et al. 2008).


Weak gravitational flexions are defined from the third derivative of gravitational lensing potential(Goldberg and Natarajan 2002,  Goldberg and Bacon 2005, Bacon, et al., 2006), hence it is more sensitive to small scale mass structure than shear which is defined as the second derivative of lensing potential.
This fact means that the flexion measurement will be very useful to measure small scale mass structure, e.g. sub-halos of galaxy clusters.

Measurement of small scale mass variation is also very important for studying mass evolution in non-linear scale by
statistical weak lensing analysis such as cosmic flexion or galaxy-galaxy flexion.
The science will be done by current and future wide field survey such as
Hyper Suprime-Cam\footnote{http://www.naoj.org/Projects/HSC/HSCProject.html} (HSC),  
Kilo-Degree Survey\footnote{http://kids.strw.leidenuniv.nl/} (KiDS),  
The Deep Lens Survey\footnote{http://matilda.physics.ucdavis.edu/working/website/index.html} (DLS),
Canada-France-Hawaii Telescope Legacy Survey\footnote{http://www.cfht.hawaii.edu/Science/CFHTLS/} (CFHTLS)
Dark Energy Survey\footnote{https://www.darkenergysurvey.org/} (DES),
and are planned such as
The Large Synoptic Survey Telescope\footnote{http://www.lsst.org} (LSST),
EUCLID\footnote{http://sci.esa.int/euclid} (EUCLID),  
Wide Field Infrared Survey Telescope\footnote{https://wfirst.gsfc.nasa.gov/} (WFIRST).
As an example, the Hyper Suprime-Cam Subaru Strategic Program (SSP) is planning 1400 deg$^2$ wide survey observation for constraining the cosmological parameters with less than 1$\%$ uncertainty.
The observation started in 2014, and  100 deg$^2$ of HSC wide survey data was recently published.
To achieve a severe constraint on the equation of state parameter for dark energy, the HSC SSP requires a highly precise weak gravitational lensing shear analysis method with lower than $1\%$ systematic error.

The flexion measurement using such a huge and detailed date is highly expected to {\bf extract} important sciences hidden in the data.
However the actual analysis of flexion measurement is very difficult because of the smallness of the flexion signal as well as complicated intrinsic noise.
Some methods to measure the flexion have been developed (Irwin and Shmakova 2006, Irwin et al. 2007, Goldberg and Leonard 2007, Levinson 2013) and mass distribution of some regions were reconstructed from measured flexion(Leonard et al. 2009, Bird and Goldberg 2018), then the application of flexion to small scale mass distribution has studied (Bacon et al.2010).
And recently, the improvement of mass reconstruction using flexion and intrinsic flexion information has been studied(Benjamin Cain 2016, Cardone 2016, Lanusse 2016, Joseph 2020).

We already developed a method to measure flexions(Okura, Umetsu and Futamase, 2007) with PSF correction, and applied the method to reconstruct the mass distribution of the Abell 1689 galaxy cluster and revealed two peak structures in the central region.
However, our analysis considered only the spin $1$ part of flexion and did not consider the combination of flexion and shear (ellipticity).
Furthermore the noise analysis associated with the flexion measurement was insufficient.      
Thus it is desirable to develop more precise flexion measurement analysis for precise mass reconstructions and sciences.

In this paper we formulate an accurate flexion measurement method based on the technique we have developed to improve the accuracy of shear measurement(Okura and Futamase 2011, 2012, 2013, 2014, 2015, 2016 and 2018).
In particular, the ERA method of PSF(Point Spread Function) correction plays a central role in our method.
In the ERA method, PSF correction is done by re-smearing the PSF and the observed galaxy image using the re-smearing function.
In this way we can use a new PSF with a simple shape and correct the PSF effect.
The method does not use any approximation to measure shear, so it does not have the systematic error in PSF correction.
The details of the ERA method can be seen in Okura 2018.

In section 2, we explain the basics about weak lensing and the basics of ERA method for shear measurement.
In section 3, we explain the basics of flexions.
In section 4, we explain the method to measure flexions.
In section 5, we summarize our conclusion.

\section{Weak lensing and ERA Method}
In this section, we explain the basics about weak lensing and the basics about ERA method for shear measurement.
In this paper, we use complex for two-dimensional value, e.g. two dimensional coordinates, and it is notated by bold font.

\subsection{Lensing distortion}
Gravitational lensing effect deflects the path of light ray from the source object by the lensing object such as star, galaxy and a cluster of galaxies.
Let the angular position of the object on the celestial sphere be $\bbe$ without lensing and $\bth$ with lensing effect.
The two-position is related by a derivative of gravitational lensing potential $\Phi$, as follows.
\begin{eqnarray}
\label{eq:lenseq}
\bbe = \bth - \partial\Phi(\bth),
\end{eqnarray}
then the coordinate $\bbe$ is called ``source plane'' and the coordinate $\bth$ is called ``image plane''.

Let $d\bbe = \bbe - \overline\bbe$ and $d\bth = \bth -\overline \bth$ be infinitesimal angular deviation from the centroid of the object image in the source plane $\overline\bbe$ and in the image plane $\overline\bth$, respectively.
If the object is at the position that the lensing effect is weak, the relation between the lengths can be described as
\begin{eqnarray}
\label{eq:dbbe}
d\bbe = \lr{1-\kappa}d\bth - \bga_L d\bth^* \equiv \lr{1-\kappa}\lr{d\bth - \mbg_L d\bth^*},
\end{eqnarray}
where $\kappa$ and $\bga_L$ are weak gravitational lensing convergence and shear defined as 2nd deviation of the lensing potential
\begin{eqnarray}
\label{eq:kappa}
\kappa&\equiv&\frac12\partial\partial^* \Phi\\
\label{eq:gamma}
\bga_L&\equiv&\frac12\partial\partial \Phi,
\end{eqnarray}
and $\mbg_L$ is reduced shear defined as $\mbg_L\equiv\bga_L/\lr{1-\kappa}$.
The equation \ref{eq:dbbe} means convergence changes the size and shear changes ellipticity of the object between source plane and image plane.

\subsection{Zero plane}
\label{seq:zeroplane}
The ERA method uses the idea of ``zero planes''.
The zero plane is an artificial plane with the coordinate $\tbbe$ where the object has a circular shape "zero image", so the zero image has zero ellipticity, then the intrinsic shape can be expressed as the distortion from the zero plane.

Let $\mbg_I$ and $\mbg_L$ be intrinsic and lensing reduced shear respectively,
then the relation between the zero and the source plane is described as
\begin{eqnarray}
d\tbbe = d\bbe - \mbg_I d\bbe^*.
\end{eqnarray}
By using equation \ref{eq:lenseq}, the relation between zero plane and image plane is described as
\begin{eqnarray}
d\tbbe &=& d\bbe - \mbg_I d\bbe^*
= \lr{1-\kappa}\lr{\lr{1+\mbg_I\mbg_L^*}d\bth - \lr{\mbg_I+\mbg_L}d\bth^*}
\nonumber\\ &\equiv &
\lr{1-\kappa}\lr{1+\mbg_I\mbg_L^*}\lr{d\bth - \mbg_C d\bth^*}\\
\mbg_C &\equiv& \frac{\mbg_I+\mbg_L}{1+\mbg_I\mbg_L^*},
\end{eqnarray}
where {\bf $\lr{1-\kappa}$ and} $\lr{1+\mbg_I\mbg_L^*}$ is just a size change and rotation, so it can be ignored in zero plane,
and $\mbg_C$ is combined with reduced shear and it is the same as the ellipticity of the object in the image plane.
Finally, by the ignoring, the lensing equation between zero plane and image plane can be described as
\begin{eqnarray}
d\tbbe = d\bth - \mbg_C d\bth^*.
\end{eqnarray}

The lensing shear $\mbg_L$ can be obtained by a simple average of the combined shear $\mbg_C$.
Let's describe the intrinsic shear distribution in polar coordinate as
\begin{eqnarray}
\mbg_I = g(r)e^{i\phi},
\end{eqnarray}
where $g(r)$ is unknown radial distribution of the intrinsic shear with normalization
\begin{eqnarray}
\int^{2\pi}_0 \hspace{-10pt} d\phi \int^1_0 \hspace{-5pt} rdr g(r) = 2\pi\int^1_0 \hspace{-5pt} rdr g(r) = 1.
\end{eqnarray}
The simple average of the combined shear with infinite number of background galaxies can be described as
\begin{eqnarray}
\label{eq:gCave}
\lrt{\mbg_C} &=& \int^{2\pi}_0 \hspace{-10pt} d\phi \int^1_0 \hspace{-5pt} rdr \frac{\mbg_I+\mbg_L}{1+\mbg_I\mbg_L^*}
\nonumber \\ &=&
\int^1_0 \hspace{-5pt} rdr \int^{2\pi}_0 \hspace{-10pt} d\phi \frac{g(r)e^{i\phi}+\mbg_L}{1+g(r)e^{i\phi}\mbg_L^*}
\nonumber \\ &=&
0+\mbg_L\int^1_0 \hspace{-5pt} rdr 2\pi g(r) = \mbg_L.
\end{eqnarray}

The notations used here after this paper are listed in table \ref{table:notations}.
\begin{table}
\begin{tabular}{c|l}\hline
notation  & definition
\\\hline\hline
$\tbbe$  &Coordinate in zero plane, the basis is the centroid of the object.
\\
$\bbe$  &Coordinate in source plane, the basis is centroid of object.
\\
$\bth$  &Coordinate in image plane, the basis is centroid of object.
\\
$\mbg$  & Weak gravitational lensing reduced shear.
\\
$\mbcF$  & Weak gravitational lensing reduced first flexion.
\\
$\mbcG$  & Weak gravitational lensing reduced second flexion.
\\
$\mbcH$  & Weak gravitational lensing reduced third flexion.
\\
$\mbcF_L$ & subscript L means Lensing distortion.
\\
$\mbcF_I$ & subscript I means Intrinsic distortion.
\\
$\mbcF_C$ & subscript C means Combined distortion.
\\
$\mbcFEB$ & subscript E means Eigen flexion.
\\
$\mbcFEC$ & subscript E' means Eigen flexion combination.
\\
$A$ &  Arbitrary object image.
\\
$W$ & Weight function.
\\
$M$ & Mask function.
\\
$J$ & Jacobian.
\\
$\cM^N_M$ & zero moment. N is order and M is spin-number.
\\
\hline
\end{tabular}
\caption{
\label{table:notations}
Notations used after next sections. These notations about image and functions are in the case of image plane, these functions in zero plane are notated with tilde, i.e. $A$ is in image plane and $\tilde A$ is in zero plane, then these functions in Fourier space is notated with hat, i.e. $\hat A$ in Fourier space.
The subscription of $\mbcF_L$, $\mbcF_I$, $\mbcF_C$, $\mbcFEB$ and $\mbcFEC$ are used for shear second flexion and third flexion with same meaning.}
\end{table}

\subsection{Zero moments and shapes}
\label{sec:Zero moments and shapes}
ERA method defines complex image moments $\cM^N_M$ of object image $A(\bth) = \tilde A(\tbbe)$ in ``zero plane'' as
\begin{eqnarray}
\cM^N_M &\equiv& \int d^2\tilde\beta d\tbbe^M_N \tilde A(\tbbe) \tilde W (|d\tbbe|) = \int d^2\theta J(\bth) \Bigl(d\tbbe(\bth)\Bigr)^M_N A(\bth) W(d\bth)\\
d\tbbe^M_N&=&d\tbbe^\frac{M+N}{2}d\tbbe^{*\frac{M-N}{2}}\\
J(\bth) &\equiv& \lrv{\frac{d^2\tilde\beta}{d^2\theta}},
\end{eqnarray}
where $\tilde W(|d\tbbe|)$ is a weight function for reducing noise from random count and $J(\bth)$ is Jacobian.
Since the object image is circular in the zero plane, the image moment with non-zero spin number must have 0 value,
so the moment is called ``Zero moment''.

The shapes of the object image, centroid and ellipticity, can be measured by finding the centroid of object $\overline\bth$ and combined reduced shear $\mbg$ which makes some zero moments zero, so
\begin{eqnarray}
\cM^1_1 &=& \int d^2\tilde\beta d\tbbe^1_1 \tilde A(\tbbe) \tilde W(|d\tbbe|) = \frac{1}{1-\lrv{\mbg_C}^2}\int d^2\theta \Bigl(d\bth - \mbg_C d\bth^*) A(\bth) W(d\bth) = 0\\
\cM^2_2 &=& \int d^2\tilde\beta d\tbbe^2_2 \tilde A(\tbbe) \tilde W(|d\tbbe|) = \frac{1}{1-\lrv{\mbg_C}^2}\int d^2\theta \Bigl(d\bth - \mbg_C d\bth^*)^2_2 A(\bth) W(d\bth) =0,
\end{eqnarray}
where $\tilde W(|d\tbbe|) = W(d\bth)$ is a weight function for reducing error in shape measurement from pixel noise on and around the galaxy image, and it must be a circular function in zero plane.

In real analysis, we should use additional mask for the weight function to
conceal unnecessary peaks appearing in the outer region of the object.
We explain about this mask  in the appendix \ref{sec:Mask}.

\section{Weak Gravitational Flexions}
In this section, we explain the basics of weak gravitational flexions.

\subsection{The basics of weak gravitational flexions}
{\bf
In the limit of weak lensing, the dominant distortion by the lensing effects is to change the size and ellipticity of the image of the background object, and the distortion can be described as the second of order deviation of the lensing potential.
However, in the region where the lensing effect is much larger, the distortion is more complex, for example it makes an arc-like shape.
In such regions, the higher-order deviation of the lensing potential must be taken into account.
}

The higher-order lensing effect for infinitesimal length is obtained by expanding lensing equation \ref{eq:lenseq} until the third deviation of lens potential, so the lensing equation with flexion is described as
\begin{eqnarray}
\label{eq:dbbe_f}
d\bbe &=&
\lr{1-\kappa}d\bth - \bga_L d\bth^* - \frac14\lr{2\mbF_L d\bth^2_0+\mbF_L^* d\bth^2_2+\mbG_L d\bth^2_{-2}}
\nonumber\\&\equiv&
\lr{1-\kappa}\lr{d\bth - \mbg_L d\bth^* - \frac14\lr{2\mbcF_L d\bth^2_0+\mbcF_L^* d\bth^2_2+\mbcG_L d\bth^2_{-2}}},
\end{eqnarray}
where $\mbF_L$ and $\mbG_L$ are first flexion and second flexion respectively, these are defined as the third deviation of lensing potential as
\begin{eqnarray}
\mbF_L &\equiv& \partial\partial\partial^*\Phi\\
\mbG_L &\equiv& \partial\partial\partial\Phi.
\end{eqnarray}
and $\mbcF_L=\mbF_L/(1-\kappa)$ and $\mbcG_L=\mbG_L/(1-\kappa)$ are first and second reduced flexion.

\subsection{Intrinsic Flexion and Combined Flexions}
\label{sec:Intrinsic Shape and Combined Flexions}
In this section, we define the combined flexions by  intrinsic flexions, namely, the first flexion and second flexion.
The definition will be made in the same way as section \ref{seq:zeroplane}.

Let the distortions(shear, first flexion and second flexion) with subscript $I$ mean intrinsic distortion, then the relations between the zero plane and the source plane is described as
\begin{eqnarray}
\label{eq:dtbbe_f}
d\tbbe &=& d\bbe - \mbg_I d\bbe^* - \frac14\lr{2\mbcF_I d\bbe^2_0+\mbcF_I^* d\bbe^2_2+\mbcG_I d\bbe^2_{-2}}.
\end{eqnarray}
Using equation \ref{eq:dbbe_f} and \label{eq:dtbbe_f}, the relation between the image plane and the zero plane with flexion are described as
\begin{eqnarray}
\label{eq:lenseq_flexion}
d\tbbe &=& \lr{1-\kappa}\lr{1+\mbg_I\mbg_L^*}\lr{d\bth - \mbg_Cd\bth^* - \frac14\lr{ 2\mbcF_C{d\bth}^2_0 +\mbcH_C{d\bth}^2_2 +\mbcG_C{d\bth}^2_{-2}}}
\end{eqnarray}
where the combined flexions $\mbcF_C, \mbcG_C, \mbcH_C$ are defined as follows
\begin{eqnarray}
\label{eq:lenseq_flexion1}
\mbcF_C&\equiv&\frac{\mbcF_L+\mbcF_I-\mbg_I\mbcF_L^*+|\mbg_L|^2\mbcF_I-\mbg_L\mbcF_I^*-\mbg_L^*\mbcG_I}{1+\mbg_I\mbg_L^*}\\
\label{eq:lenseq_flexion2}
\mbcG_C&\equiv&\frac{\mbcG_L+\mbcG_I-\mbg_I\mbcF_L-2\mbg_L\mbcF_I+\mbg_L^2\mbcF_I^*}{1+\mbg_I\mbg_L^*}\\
\label{eq:lenseq_flexion3}
\mbcH_C&\equiv&\frac{\mbcF_L^*+\mbcF_I^*-\mbg_I\mbcG_L^*-2\mbg_L^*\mbcF_I+\mbg_L^{*2}\mbcG_I}{1+\mbg_I\mbg_L^*},
\end{eqnarray}
where we call $\mbcH_C$ as the third flexion in this paper,
and here after $\lr{1-\kappa}\lr{1+\mbg\mbg_L^*}$ is ignored for the same reason explained in section \ref{seq:zeroplane}.

The above equations mean that the conjugate of $\mbcH_C$ has spin 1 and is similar as $\mbcF_C$, but the differences from $\mbcF_C$ contain variables such as $\mbg_L$(lensing distortion) and $\mbg_I$(intrinsic distortion) which are not obtained from the measurement of the individual images.
Thus these 3 quantities are independent. 
{\bf
This makes a serious problem.
The shapes of the object image $A(\bth)$ with flexion distortion are measured by requiring their corresponding zero moments to zero, where the zero moment $\cM^N_M$ is measured as follows
\begin{eqnarray}
\label{eq:zeroM_flexion}
\cM^N_M &=& \int d^2\tilde\beta d\tbbe^N_M \tilde A(\tbbe) \tilde W(|d\tbbe|) = \int d^2\theta J(\bth)\lr{d\bth - \mbg_Cd\bth^* - \frac14\lr{ 2\mbcF_C{d\bth}^2_0 +\mbcH_C{d\bth}^2_2 +\mbcG_C{d\bth}^2_{-2}}}^N_M A(\bth) W(d\bth) = 0.
\end{eqnarray}
This equation means there are five shape parameters, i.e. $\overline \bth$, $\mbg_C$, $\mbcF_C$, $\mbcG_C$ and $\mbcH_C$, to specify the zero moment.
On the other hand, there are four zero moments which are highly co-related with the five shapes, i.e. $\cM^1_1$, $\cM^2_2$, $\cM^3_1$ and $\cM^3_3$.
Although we have to find the five shape parameters which make the four zero moments zero, it is not possible to determine the five shape parameters uniquely, so one constraint condition is needed to determine them uniquely.
In the five shape parameters, centroid, shear and flexion have different properties, so the constraint condition should be imposed to relate remaining three flexions.
However, it is conceivable that the general condition may depend on the intrinsic and the lensing distortions which cannot be obtained in shape measurement of each object.
}
We need to find a particular condition in such a way that the resultant flexions do not produce any systematic error.

\begin{figure}
 \begin{minipage}{0.5\hsize}
  \begin{center}
   \includegraphics[width=70mm]{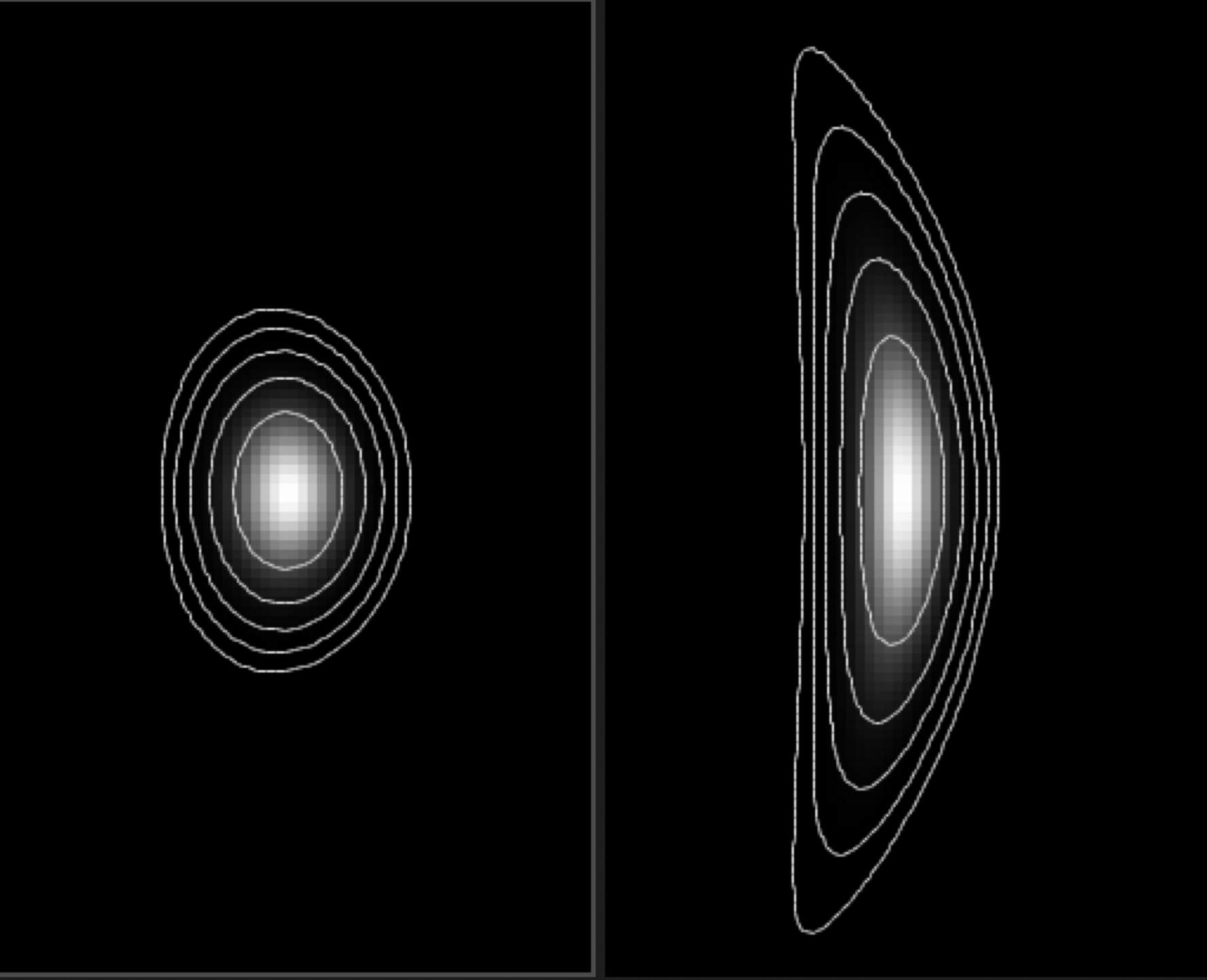}
  \end{center}
  \caption{
  \label{fig:SimLensImage}
  Gaussian images with the lensing distortion which is used in the simulation test in this paper.
{\bf
The left side image, weak arc(WA), is distorted by lensing distortion $\mbg=-0.3$, $\mbcF=-0.005$/pixels and $\mbcG=0.005$/pixels, and the right image, strong arc(SA), is distorted by lensing distortion $\mbg=-0.6$, $\mbcF=-0.01$/pixels and $\mbcG=0.01$/pixels. 
}
The brightness of the image is linear scale, but the contour is log scale for visibility of arc shape.}
 \end{minipage}
\hspace{3mm}
 \begin{minipage}{0.5\hsize}
  \begin{center}
   \includegraphics[width=70mm]{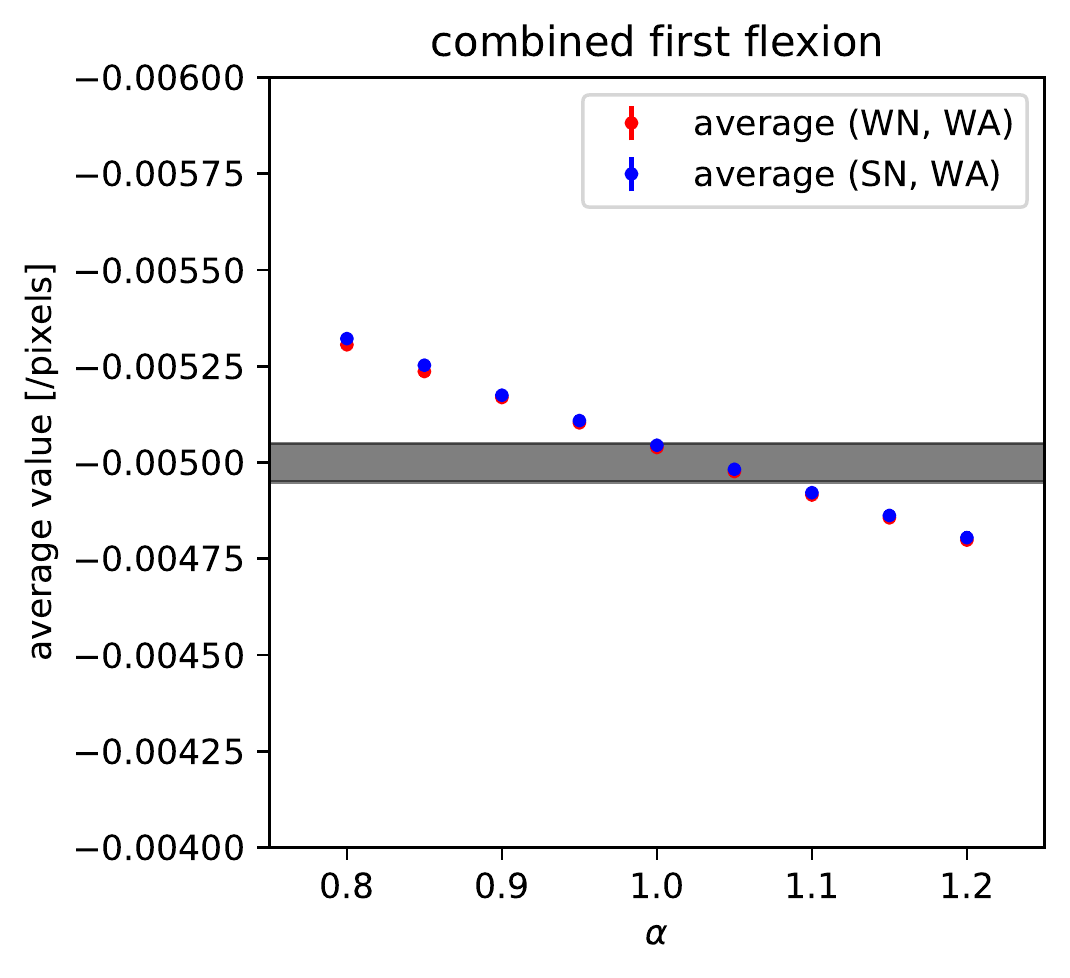}
  \end{center}
  \caption{The average value of the combined first flexion in the simulation test with the first constraint condition.
The horizontal axis is the parameter of the constraint condition $\alpha$.
The red and blue points mean the result of the data set (WN, WA) and (SN, WA) respectively; the detail of the set is described in the main text.
{\bf The gray region means true flexion value with +/- 1\% error.}
We can see the average values are strongly dependent on the constraint condition.}
  \label{fig:CFF1}
 \end{minipage}
\end{figure}

We calculate the value of the ensemble average of the three combined flexions using a simple simulated image and two constraint conditions.
In this simulation we used a simple Gaussian image which has 4 pixels Gaussian radius.
We make a lensed image by applying intrinsic and lensing distortions to the Gaussian image.
{\bf 
We used two standard deviation sets as intrinsic distortions weak noise(WN) and strong noise(SN).
The WN has  noise with $\sigma_\mbg=0.2$, $\sigma_\mbcF=0.001$/pixels(0.004 when convert to non-dimensional flexion NF with the Gaussian radius 4.0 pixels) and $\sigma_\mbcG=0.001$/pixels(NF=0.004),
and  the SN has noise with $\sigma_\mbg=0.3$, $\sigma_\mbcF=0.005$/pixels(NF=0.02) and $\sigma_\mbcG=0.005$/pixels(NF=0.02).
Then we used two lensing distortions, weak arc(WA) and strong arc(SA).
The WA has distortion $\mbg_L=-0.3$, $\mbcF_L=-0.005$/pixels(NF=-0.02) and $\mbcG_L=0.005$/pixels(NF=0.02),
and The SA has distortion $\mbg_L=-0.6$, $\mbcF_L=-0.01$/pixels(NF=-0.04) and $\mbcG_L=0.01$/pixels(NF=0.04).
Figure \ref{fig:SimLensImage} shows sample images which are Gaussian images distorted by only the lensing distortions, so this test investigates also the precision of flexion measurement from such an arc shape object.
}

\begin{figure}
 \begin{minipage}{0.495\hsize}
  \begin{center}
\vspace{-4mm}
   \includegraphics[width=70mm]{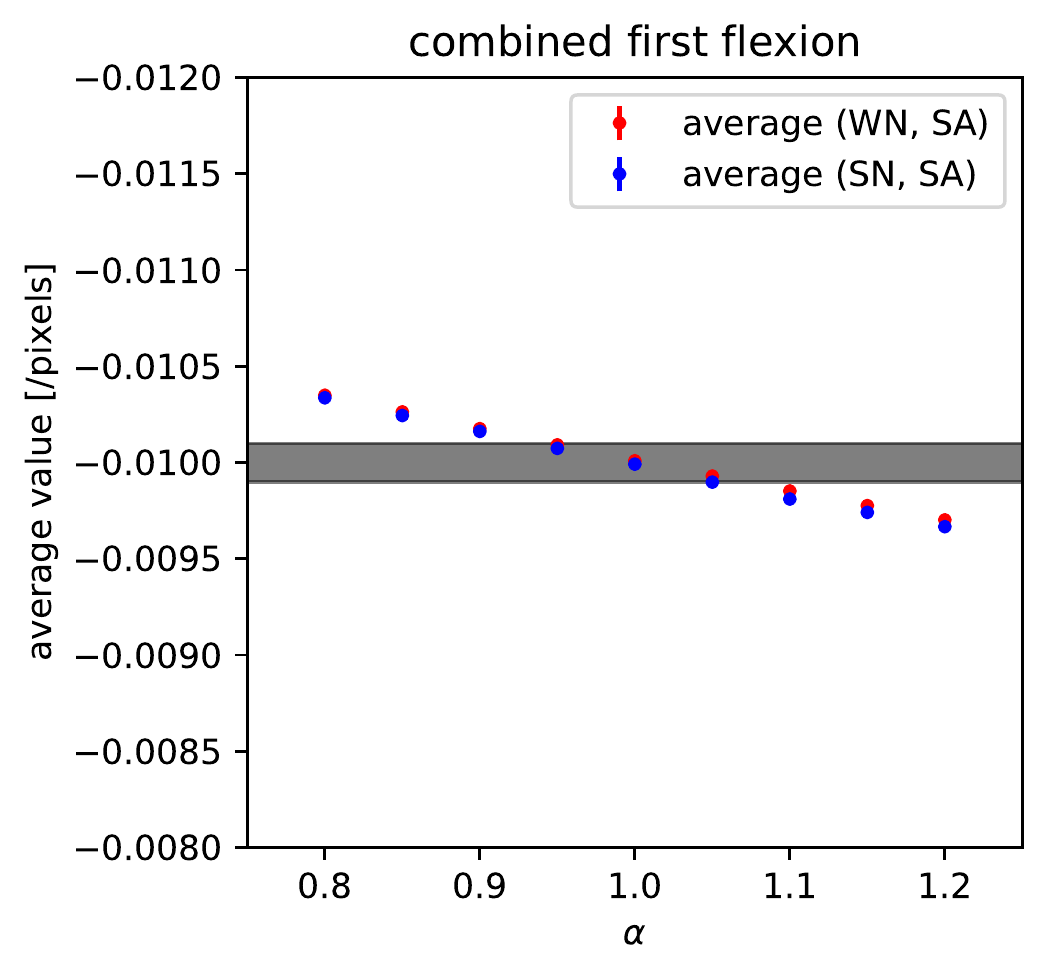}
  \end{center}
  \caption{
  Same plot as figure \ref{fig:CFF1} except that the red and blue points are in the case of (WN, SA) and (SN, SA) respectively.}
  \label{fig:CFF2}
 \end{minipage}
\hspace{3mm}
 \begin{minipage}{0.495\hsize}
  \begin{center}
   \includegraphics[width=70mm]{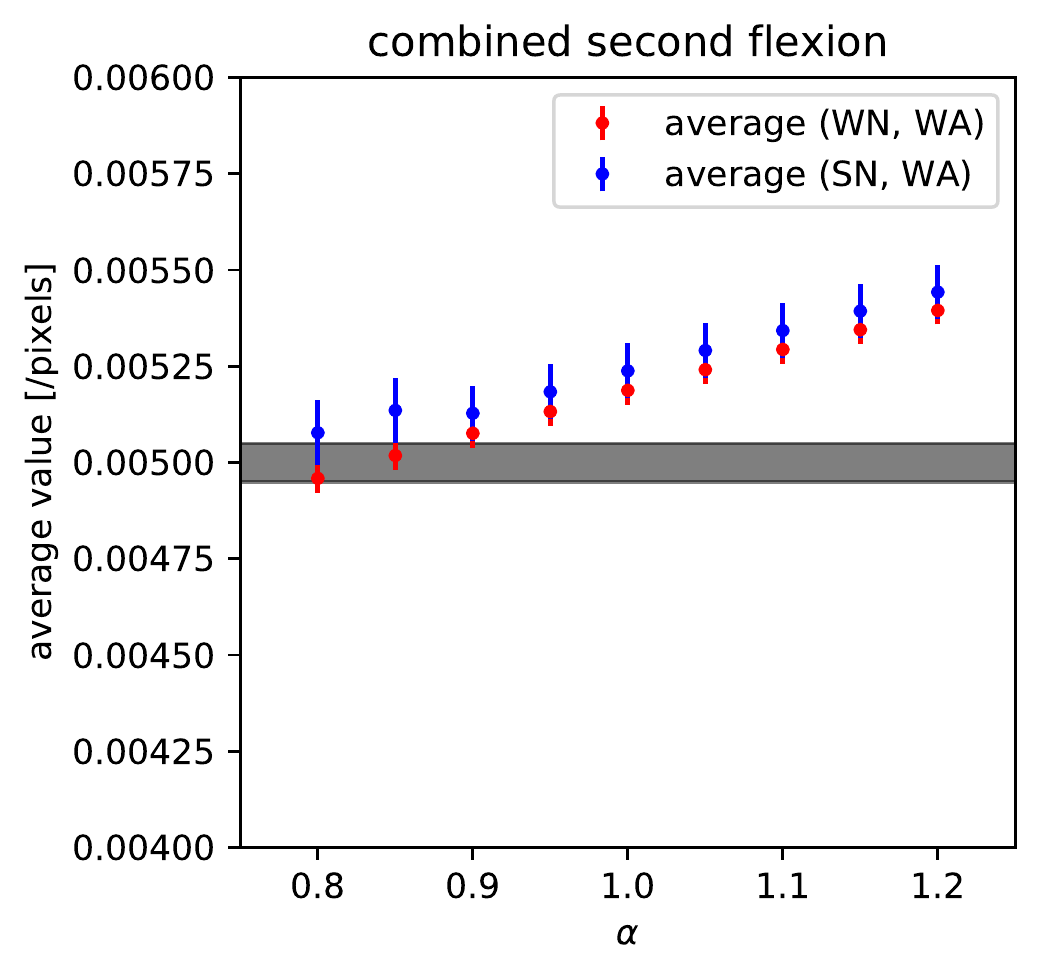}
  \end{center}
  \caption{
Same plot as figure \ref{fig:CFF1} except that the red and blue points are the average value of combined second flexion in the case of (WN, WA) and (SN, WA) respectively.}
  \label{fig:CFG1}
 \end{minipage}
\end{figure}

\begin{figure}
 \begin{minipage}{0.495\hsize}
  \begin{center}
\vspace{-4mm}
   \includegraphics[width=70mm]{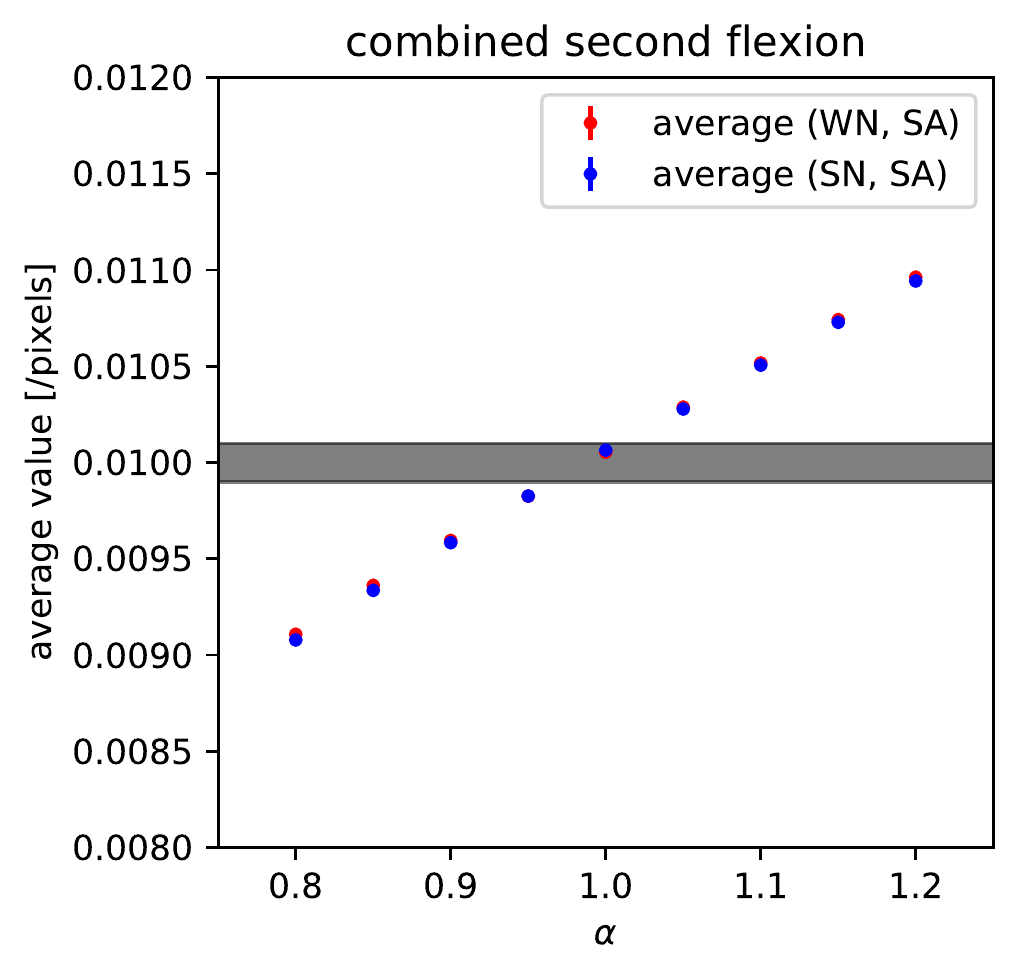}
  \end{center}
  \caption{
Same plot as figure \ref{fig:CFG1} except that the red and blue points are in the case of (WN, SA) and (SN, SA) respectively.}
  \label{fig:CFG2}
 \end{minipage}
\hspace{3mm}
 \begin{minipage}{0.495\hsize}
  \begin{center}
   \includegraphics[width=70mm]{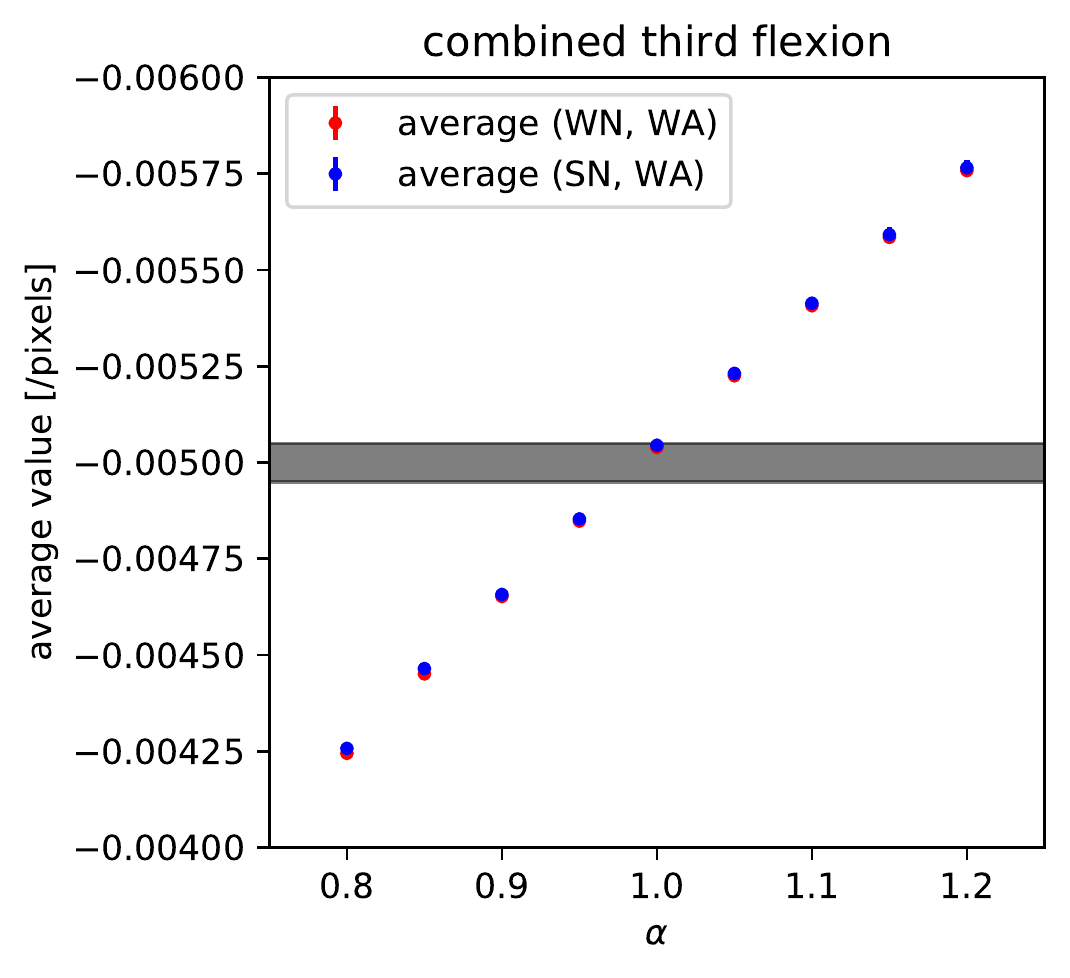}
  \end{center}
  \caption{
Same plot as figure \ref{fig:CFF1} except that the red and blue points are the average value of combined third flexion in the case of (WN, WA) and (SN, WA) respectively.}
  \label{fig:CFH1}
 \end{minipage}
\end{figure}

\begin{figure}
 \begin{minipage}{0.495\hsize}
  \begin{center}
\vspace{-14mm}
   \includegraphics[width=70mm]{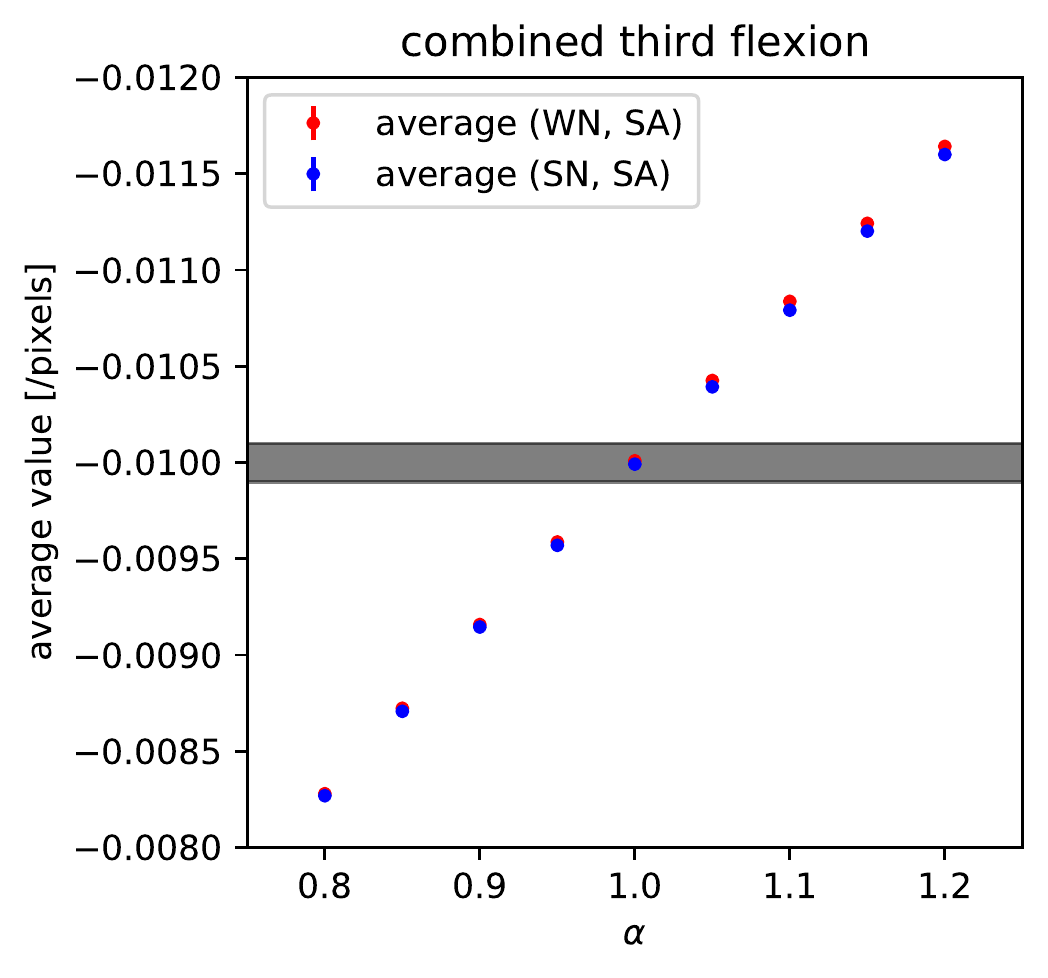}
  \end{center}
  \caption{
Same plot as figure \ref{fig:CFH1} except that the red and blue points are in the case of (WN, SA) and (SN, SA) respectively.}
  \label{fig:CFH2}
 \end{minipage}
\hspace{3mm}
 \begin{minipage}{0.495\hsize}
  \begin{center}
   \includegraphics[width=70mm]{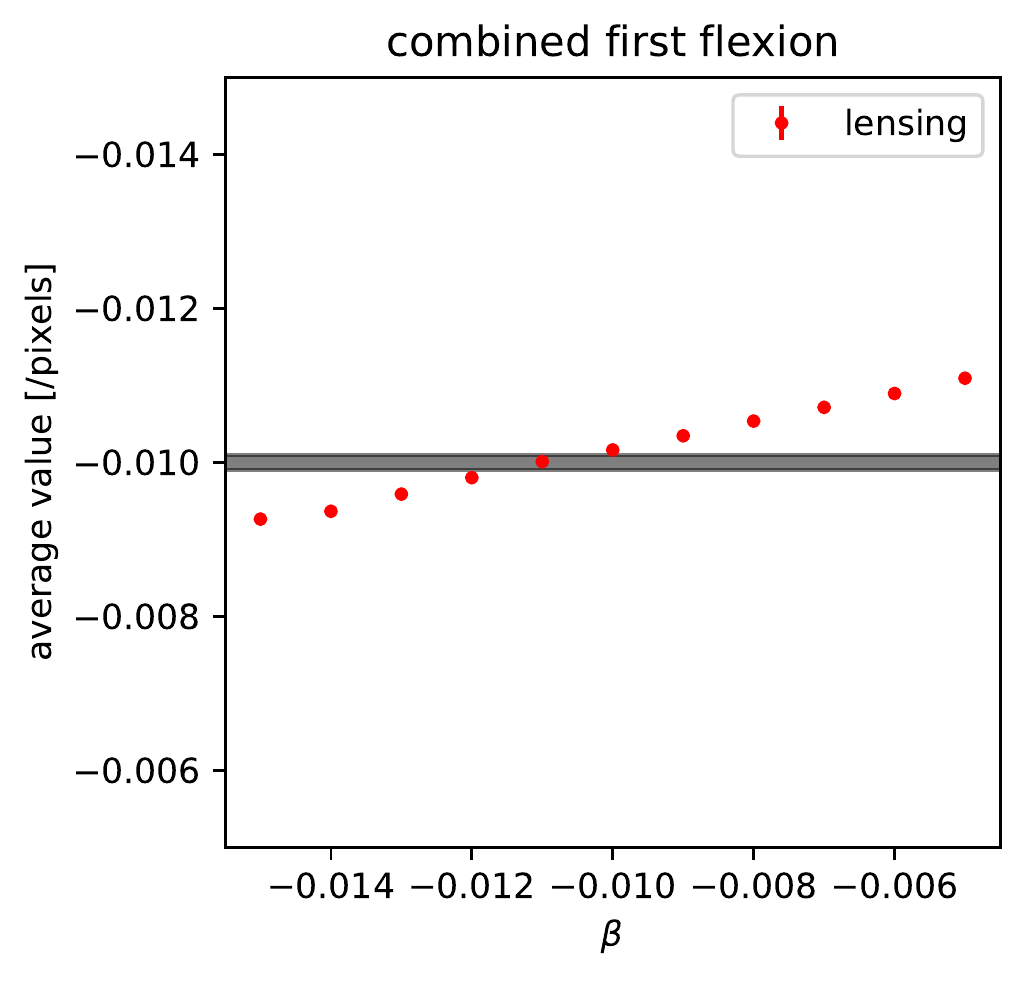}
  \end{center}
  \caption{The average value of the combined first flexion in the simulation test with the second constraint condition.
The horizontal axis is the parameter of the constraint condition $\beta$.
The red  points mean the result of the data set (SN, SA).
The solid line is combined first flexion without any intrinsic distortions, so lensing first flexion.
We can see the average values are strongly dependent on the constraint condition.}
  \label{fig:CFF1_fix}
 \end{minipage}
\end{figure}

\begin{figure}
 \begin{minipage}{0.495\hsize}
  \begin{center}
\vspace{0mm}
   \includegraphics[width=70mm]{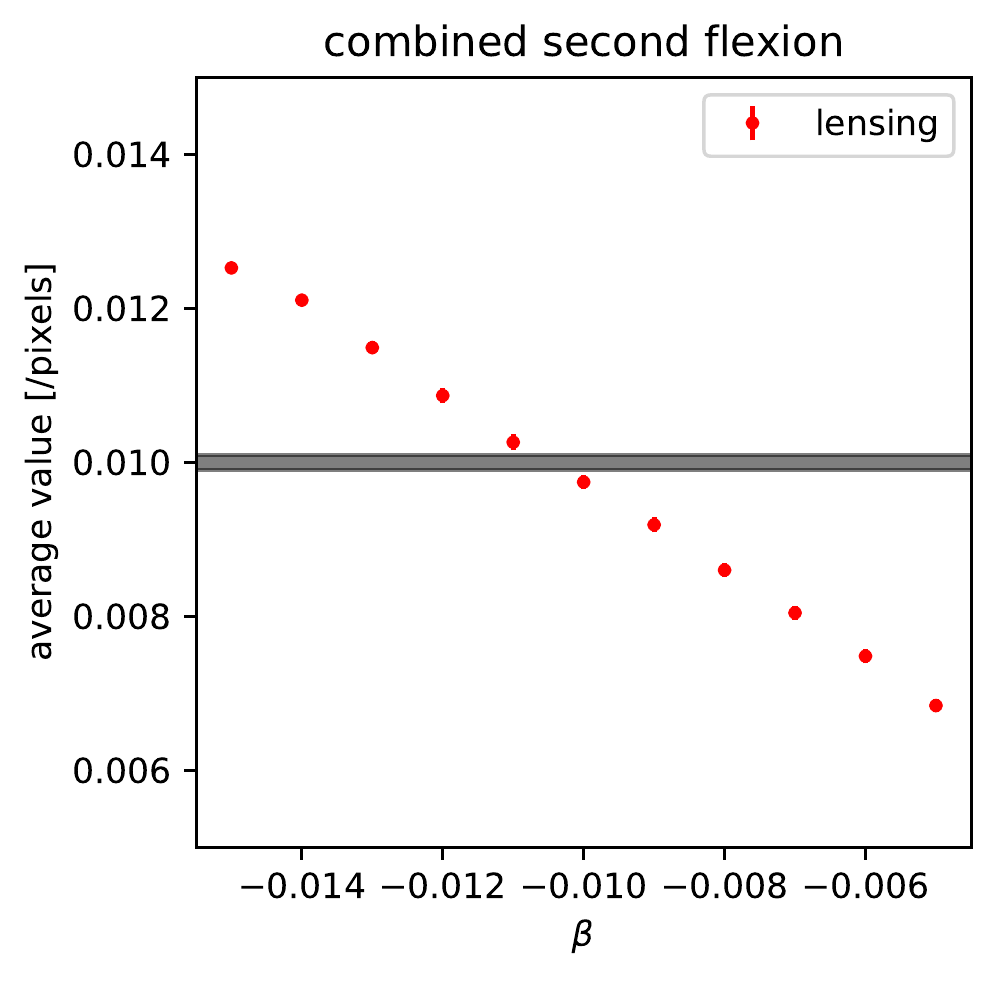}
  \end{center}
  \caption{
Same plot as figure \ref{fig:CFF1_fix} except that the red points are the average value of combined second flexion.}
  \label{fig:CFG1_fix}
 \end{minipage}
\hspace{3mm}
 \begin{minipage}{0.495\hsize}
  \begin{center}
   \includegraphics[width=70mm]{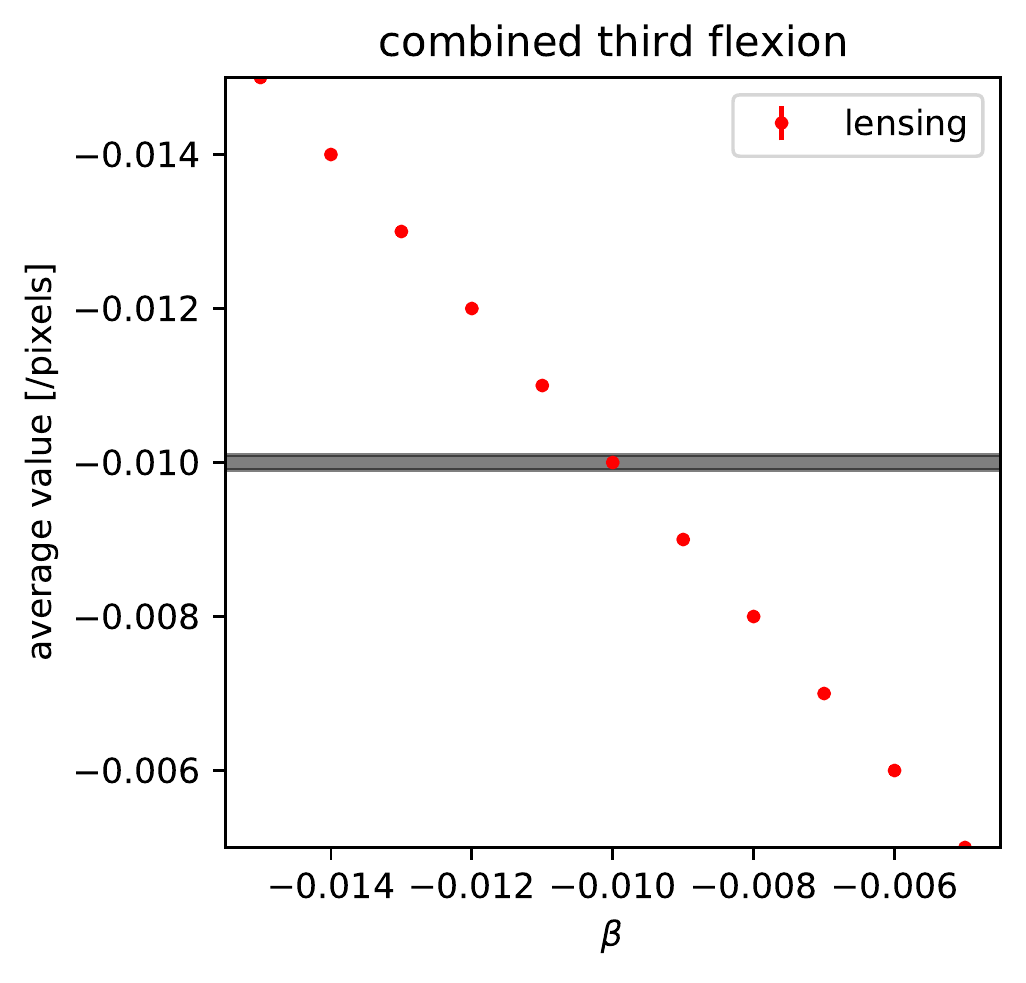}
  \end{center}
  \caption{
Same plot as figure \ref{fig:CFF1_fix} except that the red points are the average value of combined third flexion.}
  \label{fig:CFH1_fix}
 \end{minipage}
\end{figure}

A conjugate of the third flexion $\mbcH_C$ is equal to $\mbcF_C$ up to the first order,  so the first one of the constraint condition we use in this simulation is
\begin{eqnarray}
\label{eq:CC_1st}
\mbcH_C=\alpha\mbcF^*_C
\end{eqnarray}
 with variable $\alpha$.
Figure \ref{fig:CFF1} to \ref{fig:CFH2} show the ensemble average of the combined flexions for variable $\alpha$ of the constraint condition.
Then, the second one of the constraint condition is simply applying constant for $\mbcH_C$, so
\begin{eqnarray}
\label{eq:CC_2nd}
\mbcH_C=\beta
\end{eqnarray}
Figure \ref{fig:CFF1_fix} to \ref{fig:CFH1_fix} show the ensemble average with variable $\beta$ of the constraint condition only for the data set (SN, SA).
We can see the combined flexions strongly depend on not only strength of intrinsic noise but also selection of the constraint condition.
 

\section{A new method for flexion precise measurement without intrinsic noise}
\label{sec:A new method for flexion precise measurement with removing intrinsic noise}
\subsection{eigen flexion values and lensing information}
\label{sec:eigen flexion values and lensing information}
In the previous section, it was shown that systematic error occurs because of the wrong choice of the constraint condition, or the wrong choice of the quantities we measure. In fact it turns out that the latter is the case and we can find two kinds of combination of the combined flexions whose average is independent of the selection of the constraint condition.
We call these combinations as eigen flexions.

In deriving the eigen flexions, we use the ERA method where the zero image in the zero plane has circular shape with arbitrary radial profile. This means that the brightness distribution of zero image is a function with variable $|d\tbbe|^2$. Then the variable is transformed in the lens plane up to the first-order in flexion as
\begin{eqnarray}
\label{eq:observable_flexion}
|d\tbbe|^2 &=& \lr{1-\kappa}^2\lrv{1+\mbg_I\mbg_L^*}^2
\lr{\lr{1+|\mbg_C|^2}d\bth^2_0 - 2\Real{\mbg_C^*d\bth^2_2} - \frac12\lr{\Real{\mbcFEB^* d\bth^3_1}+\Real{\mbcGEB^*d\bth^3_3}}}\\
\label{eq:eigen_flexion1}
\mbcFEB &\equiv& 2\mbcF_C+\mbcH^*_C-2\mbg_C\mbcF^*_C-\mbg^*_C\mbcG_C\\
\label{eq:eigen_flexion2}
\mbcGEB &\equiv& \mbcG_C-\mbg_C\mbcH^*_C
\end{eqnarray}
Natural combinations of the combined flexions appear here. This equation means that the brightness distribution of the lensed image is a function with variables $\mbg_C$, $\mbcFEB$ and $\mbcGEB$, and the flexion effect is described by  $\mbcFEB$ and $\mbcGEB$. We call these as eigen flexions.

\begin{figure}
 \begin{minipage}{0.495\hsize}
  \begin{center}
\vspace{2mm}
   \includegraphics[width=70mm]{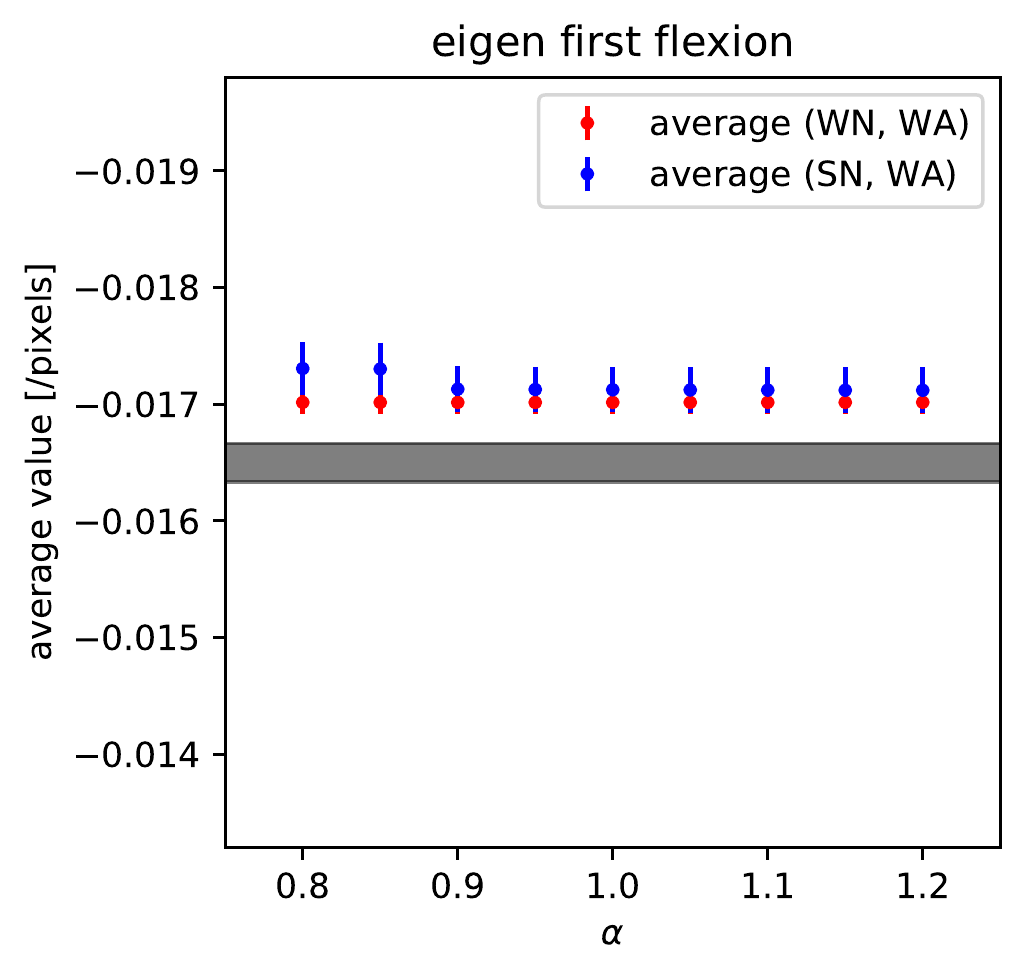}
  \end{center}
  \caption{
Same plot as figure \ref{fig:CFF1} except that the red and blue points are the average value of eigen first flexion.
We can see fluctuation of the average value decreases compare to the average of combined first flexion in all $\alpha$ region.}
  \label{fig:EFBF1}
 \end{minipage}
\hspace{3mm}
 \begin{minipage}{0.495\hsize}
  \begin{center}
   \includegraphics[width=70mm]{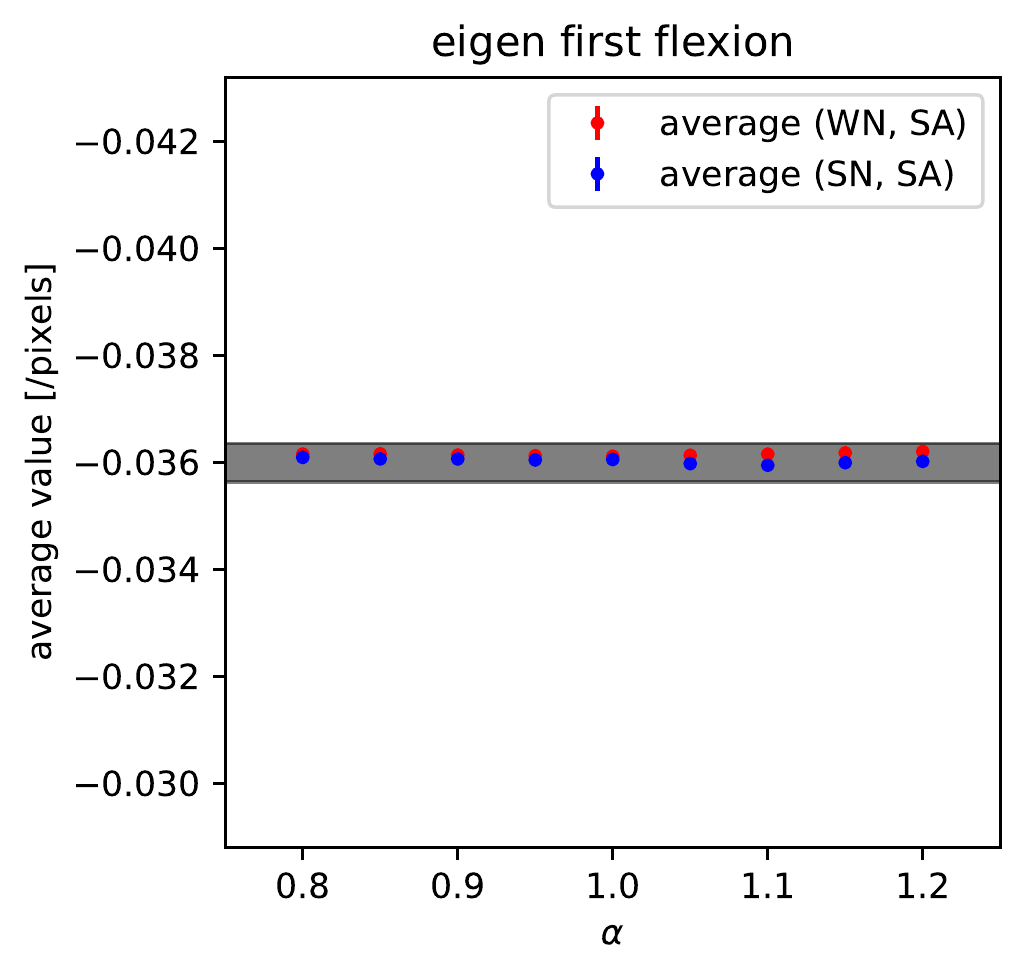}
  \end{center}
  \caption{
Same plot as figure \ref{fig:EFBF1} except that the red and blue points are in the case of (WN, SA) and (SN, SA) respectively.}
  \label{fig:EFBF2}
 \end{minipage}
\end{figure}

\begin{figure}
 \begin{minipage}{0.495\hsize}
  \begin{center}
\vspace{-0mm}
   \includegraphics[width=70mm]{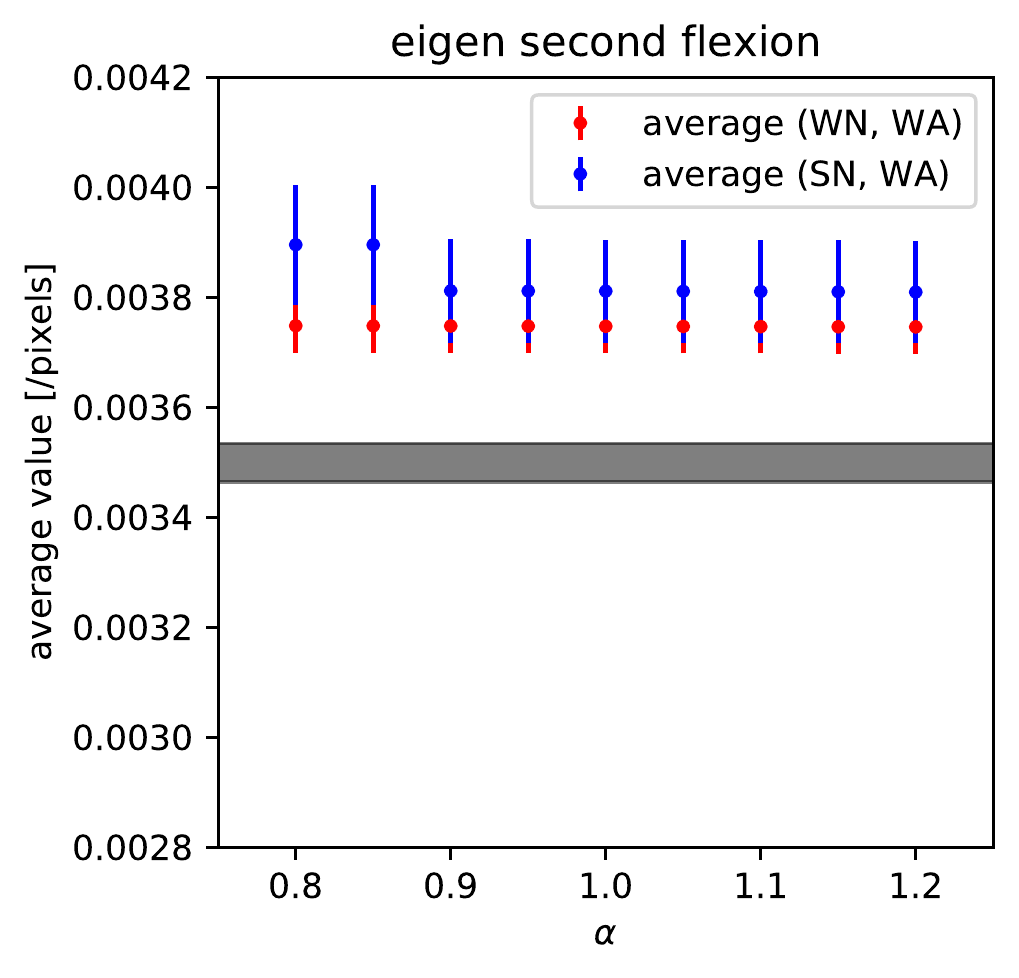}
  \end{center}
  \caption{
Same plot as figure \ref{fig:EFBF1} except that the red and blue points are the average value of eigen second flexion.}
  \label{fig:EFBG1}
 \end{minipage}
\hspace{3mm}
 \begin{minipage}{0.495\hsize}
  \begin{center}
   \includegraphics[width=70mm]{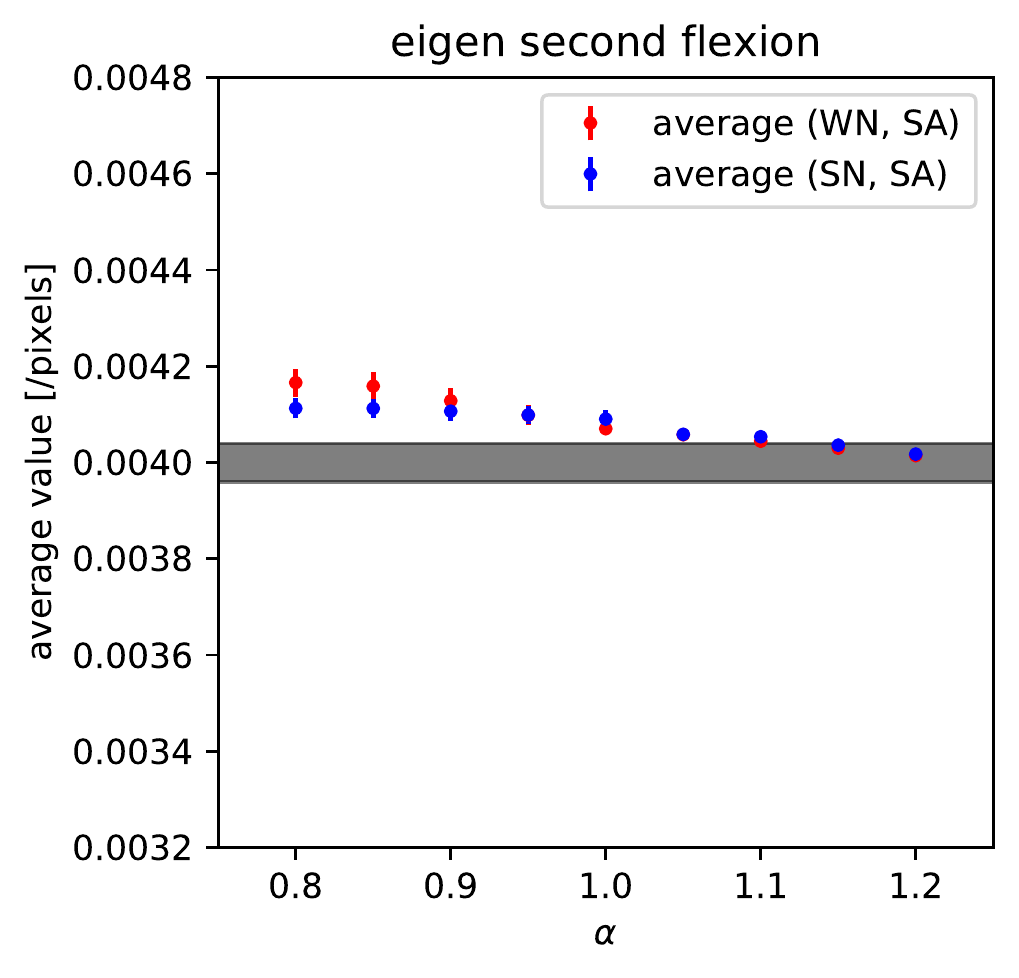}
  \end{center}
  \caption{
Same plot as figure \ref{fig:EFBG1} except that the red and blue points are in the case of (WN, SA) and (SN, SA) respectively.}
  \label{fig:EFBG2}
 \end{minipage}
\end{figure}

\begin{figure}
 \begin{minipage}{0.495\hsize}
  \begin{center}
\vspace{2mm}
   \includegraphics[width=70mm]{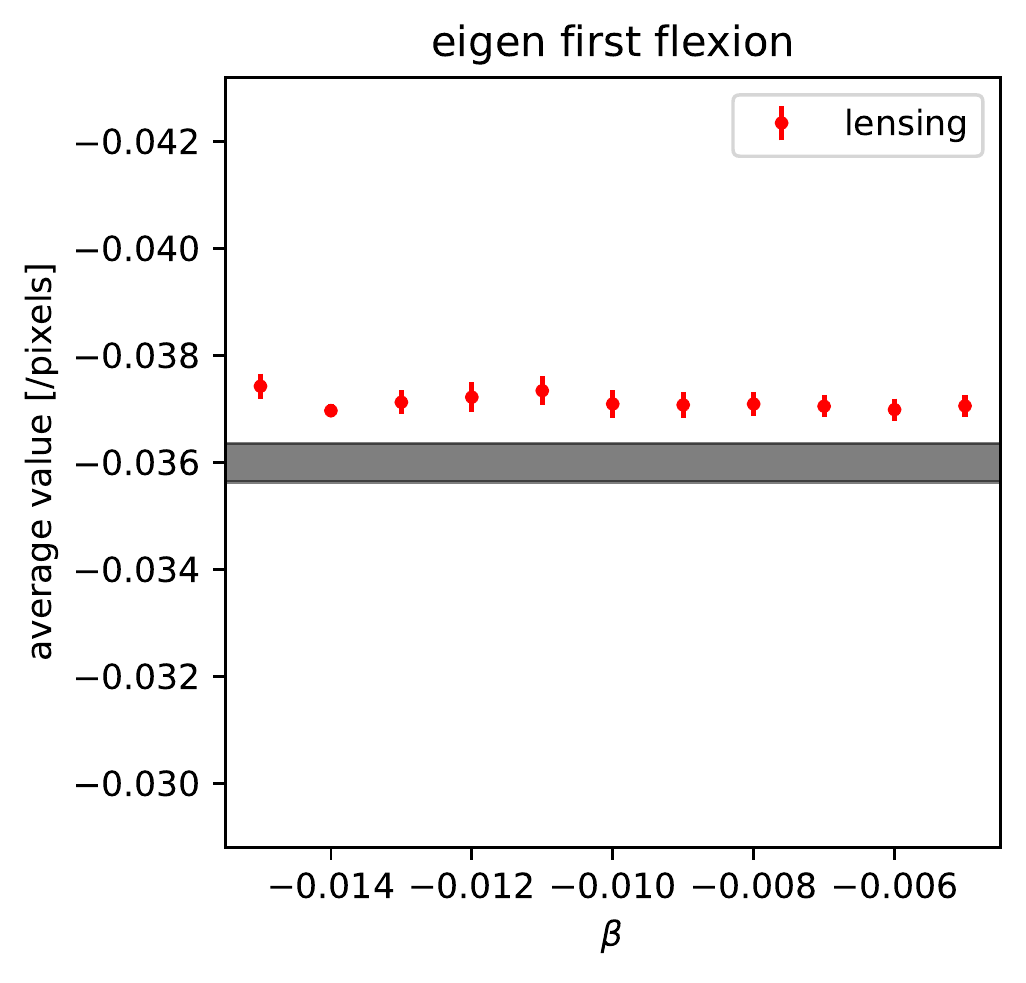}
  \end{center}
  \caption{
Same plot as figure \ref{fig:EFBF1} except using the second constraint condition, the horizontal axis means the constant constraint condition $\beta$ and points are only (SN, SA).}
  \label{fig:EFB1_fix}
 \end{minipage}
\hspace{3mm}
 \begin{minipage}{0.495\hsize}
  \begin{center}
   \includegraphics[width=70mm]{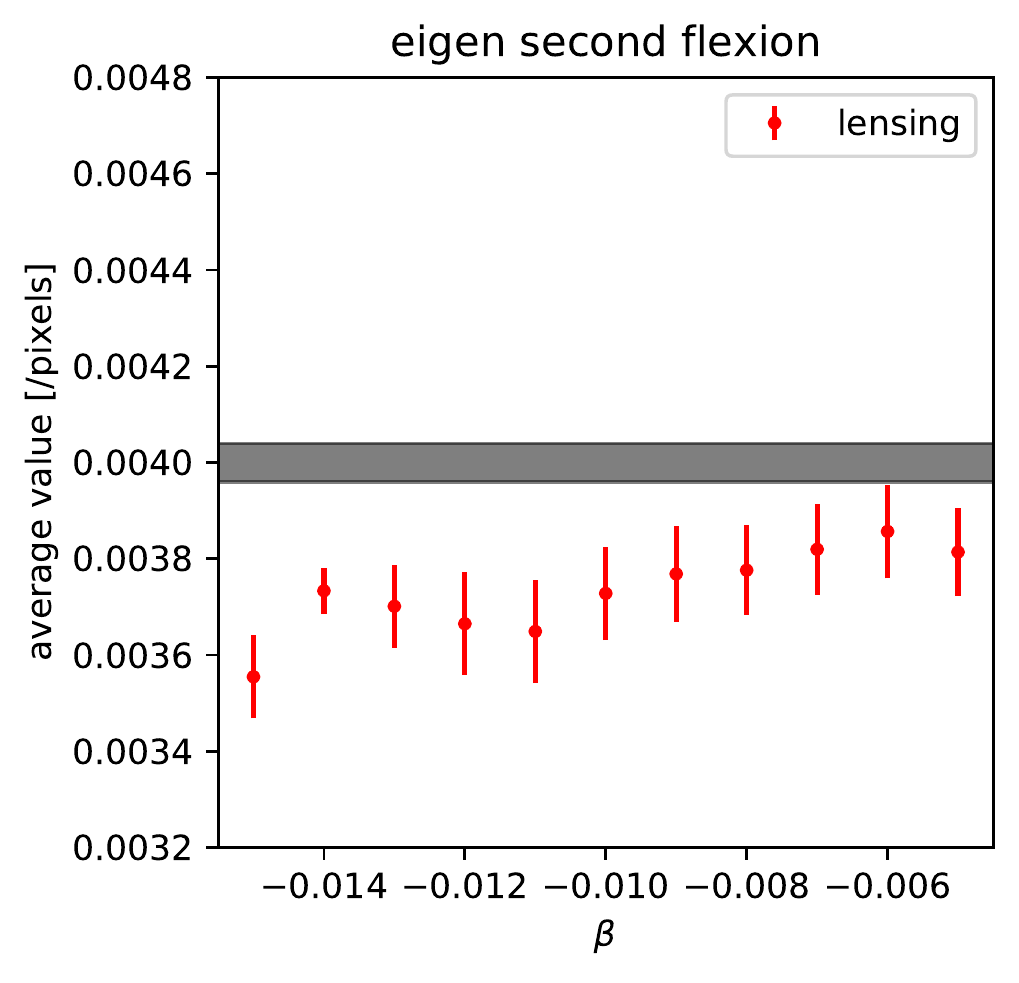}
  \end{center}
  \caption{
Same plot as figure \ref{fig:EFB1_fix} except that the red points are the average value of eigen second flexion.}
  \label{fig:EGB1_fix}
 \end{minipage}
\end{figure}

Using the simulation described in the previous section, we have confirmed the independence of the eigen flexion from the choice of the above two constraint conditions with the same simulation test as the previous section. ( Actually we have checked the independence of the result using other constraint conditions.)
Figures \ref{fig:EFBF1} to \ref{fig:EGB1_fix} are the result of the simulation test.
We can see the fluctuations of the average of the eigen flexion decrease compared to the average of the combined flexions.
So, we can obtain the eigen flexion $\mbcFEB$ and $\mbcGEB$ by using arbitrary convenient value of the parameter in the constraint condition from the combined  flexion $\mbcF_C$ and $\mbcG_C$.
{\bf
However, because the eigen flexions still contain the intrinsic shear term $|\mbg_I|^2$, the ensemble average of the eigen flexions have systematic error due to intrinsic noise.
}

We have found a combination of eigen flexion which do not include $|\mbg_I|^2$ as follows
\begin{eqnarray}
\label{eq:z_flexion1}
\mbcFEC &\equiv&\frac{\mbcFEB+\mbg^*_C\mbcGEB+\mbg_C\mbcFEB^*+\mbg_C^2\mbcGEB^*}{1-|\mbg_C|^2}=2\mbcF_C+\mbcH^*_C+\mbg_C\mbcH_C\\
\label{eq:z_flexion2}
\mbcGEC &\equiv&\frac{\mbcGEB+\mbg_C\mbcFEB+\mbg^2_C\mbcFEB^*+\mbg^3_C\mbcGEB^*}{1-|\mbg_C|^2}=\mbcG_C+2\mbg_C\mbcF_C+\mbg^2_C\mbcH_C.
\end{eqnarray}
Finally, by taking ensemble average of them, we can remove intrinsic distortions,
\begin{eqnarray}
\label{eq:ave_z_flexion1}
\lrt{\mbcFEC} &=&3\mbcF_L+\mbg_L\mbcF^*_L\\
\label{eq:ave_z_flexion2}
\lrt{\mbcGEC} &=&\mbcG_L+2\mbg_L\mbcF_L+\mbg^2_L\mbcF^*_L,
\end{eqnarray}
where we assume that the intrinsic distortions are independent of each other.
Equation \ref{eq:ave_z_flexion1} and \ref{eq:ave_z_flexion2} still contain lensing shear $\mbg_L$, but it can be obtained by ensemble average of combined shear $\mbg_C$.
We call the two flexions $\lrt{\mbcFEC}$ and $\lrt{\mbcGEC}$ as an eigen (first and second) flexion combination.

\begin{figure}
 \begin{minipage}{0.495\hsize}
  \begin{center}
\vspace{2mm}
   \includegraphics[width=70mm]{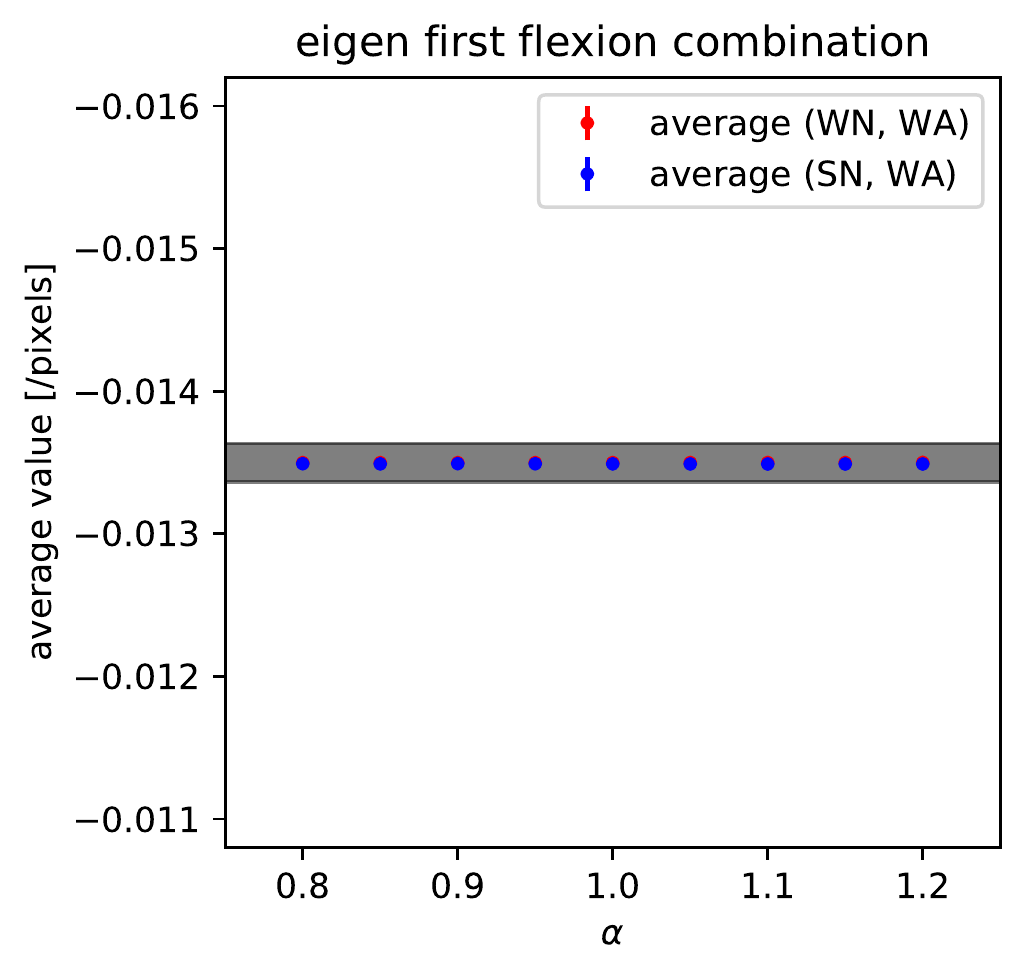}
  \end{center}
  \caption{
Same plot as figure \ref{fig:CFF1} except that the red and blue points are the average value of eigen first flexion combination.
We can see the average values have same value as lensing value in all $\alpha$ range.}
  \label{fig:EFCF1}
 \end{minipage}
\hspace{3mm}
 \begin{minipage}{0.495\hsize}
  \begin{center}
   \includegraphics[width=70mm]{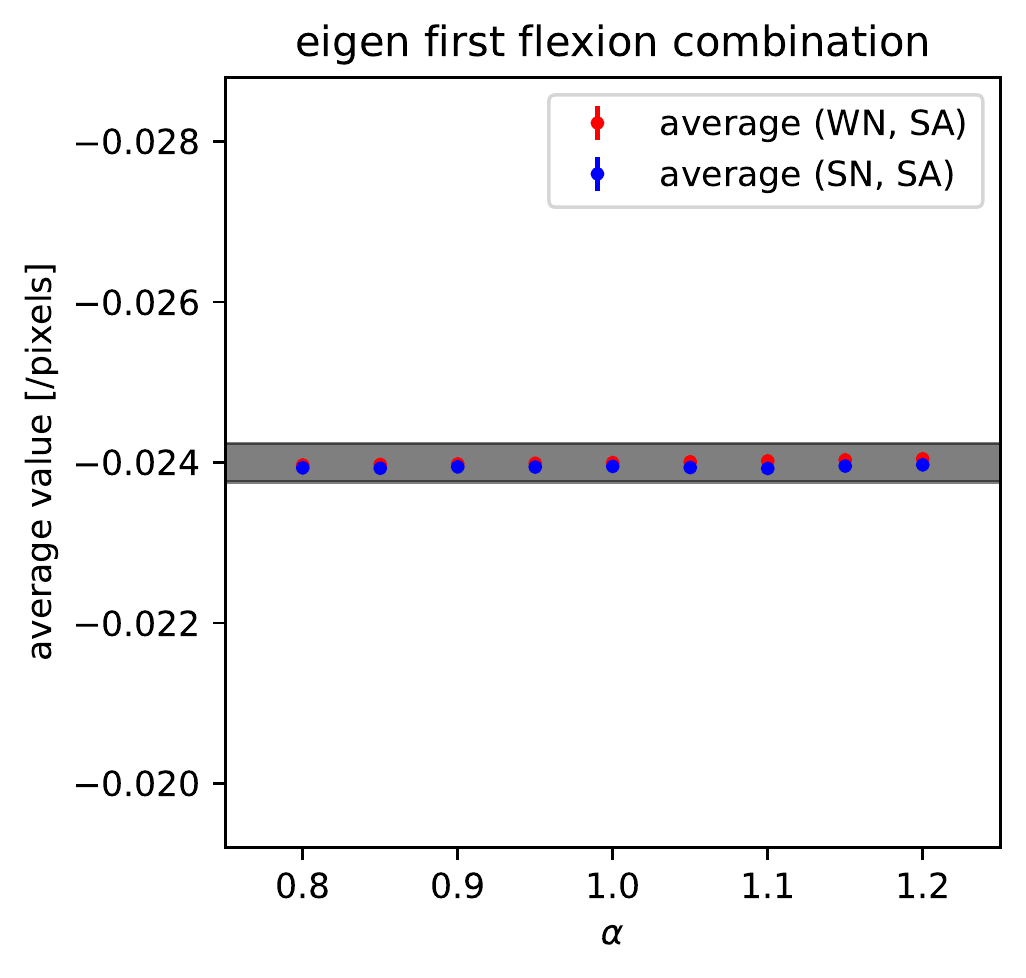}
  \end{center}
  \caption{
Same plot as figure \ref{fig:EFCF1} except that the red and blue points are in the case of (WN, SA) and (SN, SA) respectively.}
  \label{fig:EFCF2}
 \end{minipage}
\end{figure}

\begin{figure}
 \begin{minipage}{0.495\hsize}
  \begin{center}
\vspace{2mm}
   \includegraphics[width=70mm]{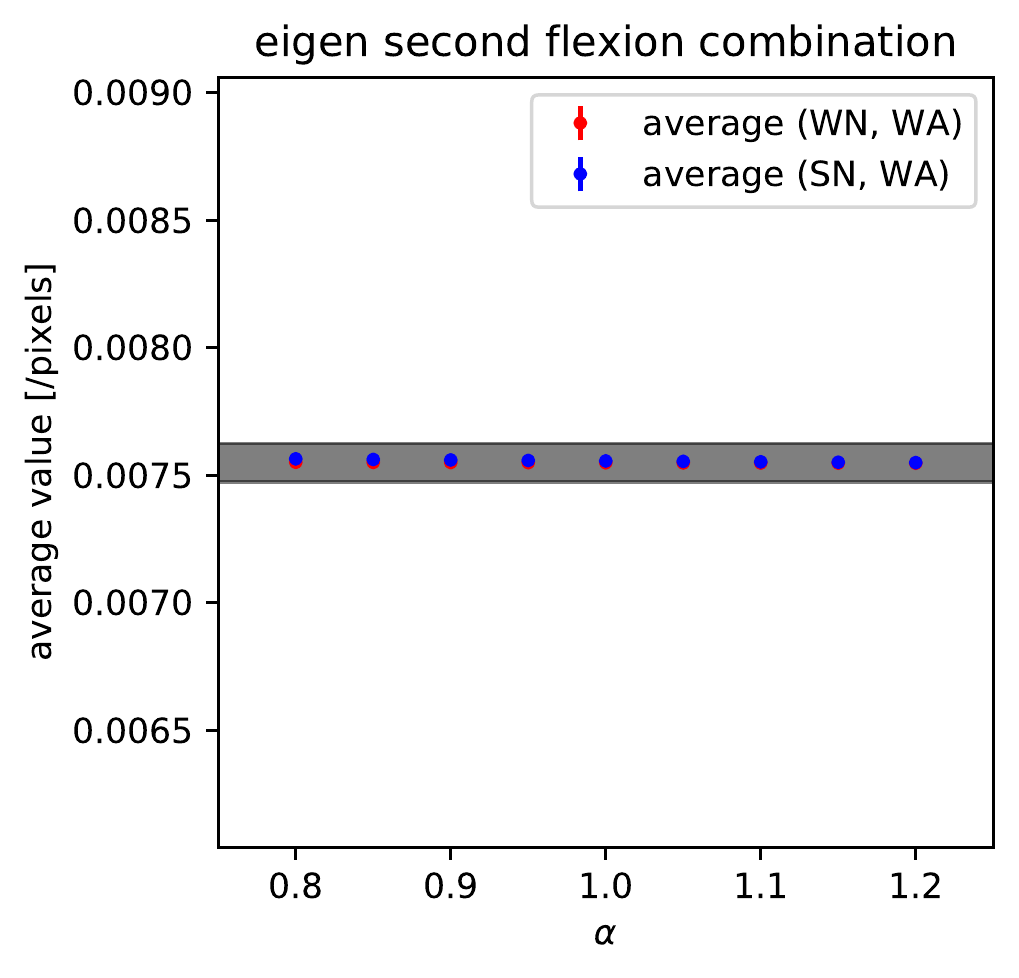}
  \end{center}
  \caption{
Same plot as figure \ref{fig:EFCF1} except that the red and blue points are the average value of eigen second flexion complex.}
  \label{fig:EFCG1}
 \end{minipage}
\hspace{3mm}
 \begin{minipage}{0.495\hsize}
  \begin{center}
   \includegraphics[width=70mm]{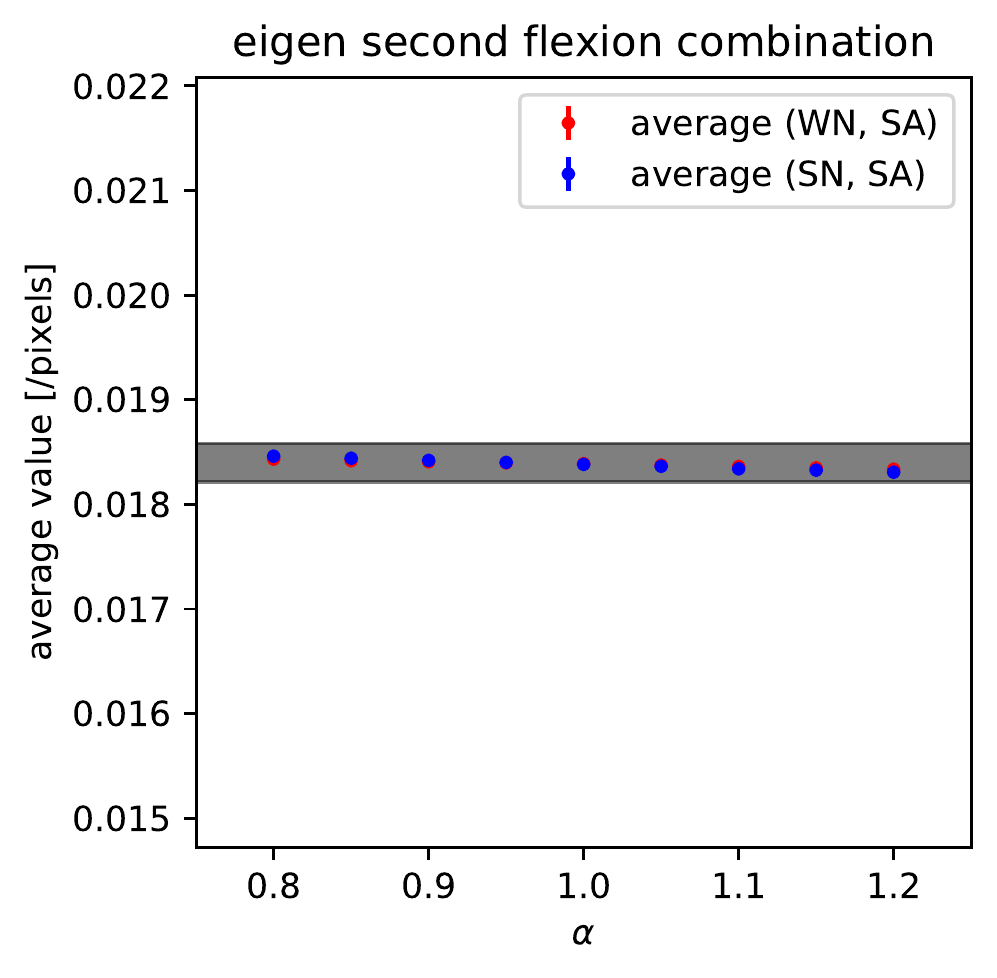}
  \end{center}
  \caption{
Same plot as figure \ref{fig:EFCG1} except that the red and blue points are the average value of eigen second flexion.}
  \label{fig:EFCG2}
 \end{minipage}
\end{figure}

\begin{figure}
 \begin{minipage}{0.495\hsize}
  \begin{center}
\vspace{2mm}
   \includegraphics[width=70mm]{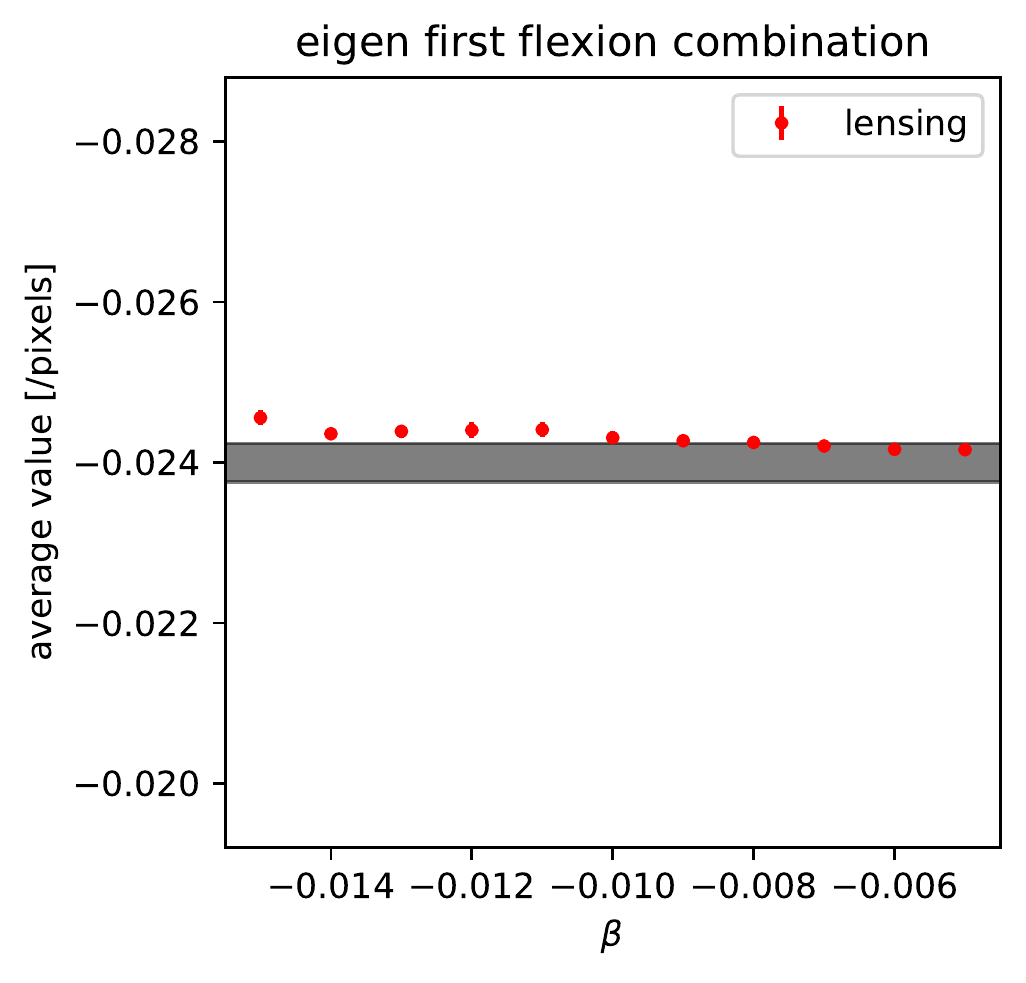}
  \end{center}
  \caption{
Same plot as figure \ref{fig:EFCF1} except using the second constraint condition, the horizontal axis means the constant constraint condition $\beta$ and points are only (SN, SA).}
  \label{fig:EFC1_fix}
 \end{minipage}
\hspace{3mm}
 \begin{minipage}{0.495\hsize}
  \begin{center}
   \includegraphics[width=70mm]{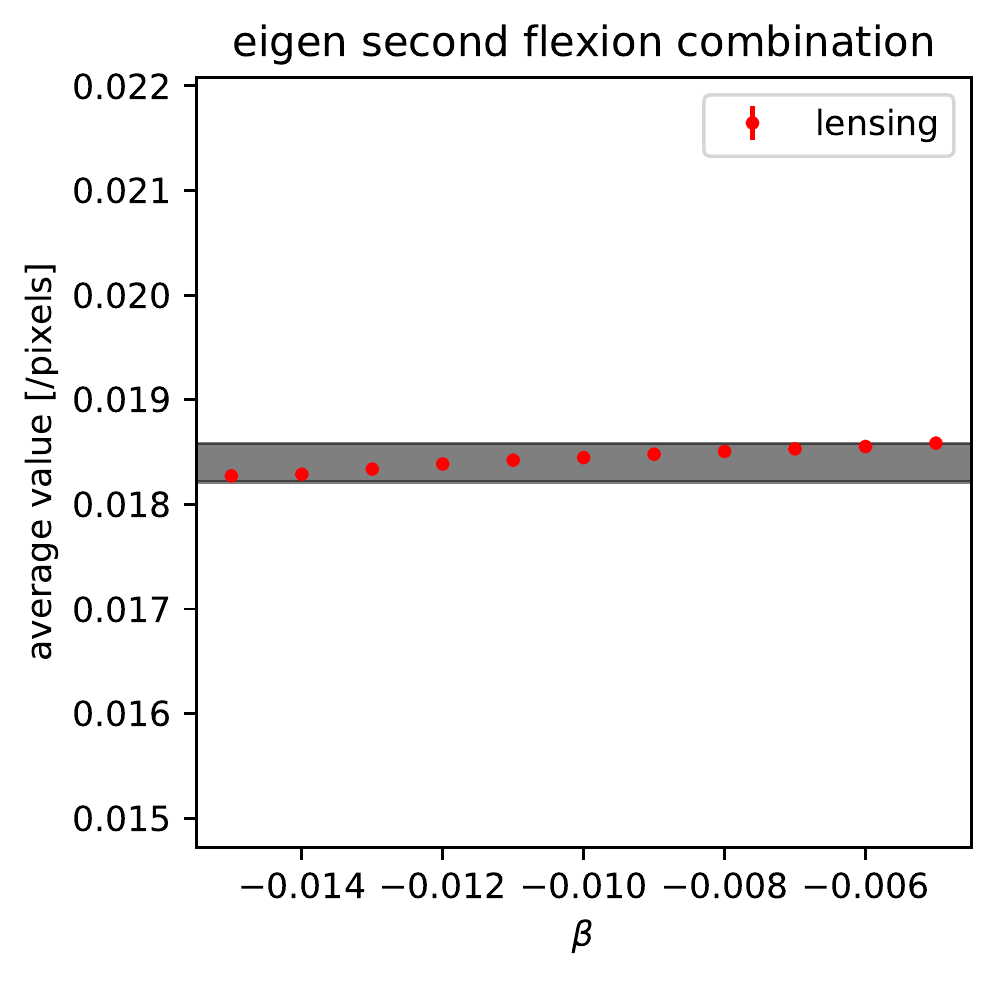}
  \end{center}
  \caption{
Same plot as figure \ref{fig:EFB1_fix} except that the red points are the average value of eigen second flexion combination.}
  \label{fig:EGC1_fix}
 \end{minipage}
\end{figure}

We investigate the eigen flexion combinations by the same simulation in the combined flexions.
Figures \ref{fig:EFCF1} to \ref{fig:EGC1_fix} are plots of the ensemble average of the eigen flexion combinations measured from the simulated images.
We can see the average value does not depend on the strength of intrinsic distortions, so the their average values have only the lensing information and coincide correctly with the given values.

It is confirmed that the ensemble average of the eigen flexion combinations have only lensing information, so finally, after obtaining $\mbg_L$ from $\lrt{\mbg_C}$, we can obtain weak gravitational lensing flexions as
\begin{eqnarray}
\label{eq:WLF1}
\mbcF_L &=&\frac{3\lrt{\mbcFEC}-\lrt{\mbg_C}\lrt{\mbcFEC}^*}{9-|\lrt{\mbg_C}|^2}\\
\label{eq:WLF2}
\mbcG_L &=&\lrt{\mbcGEC}-\lrt{\mbg_C}\frac{\lr{6-|\lrt{\mbg_C}|^2}\lrt{\mbcFEC}+\lrt{\mbg_C}\lrt{\mbcFEC}^*}{9-|\lrt{\mbg_C}|^2},
\end{eqnarray}

\begin{figure*}
\centering
\resizebox{0.95\hsize}{!}{\includegraphics{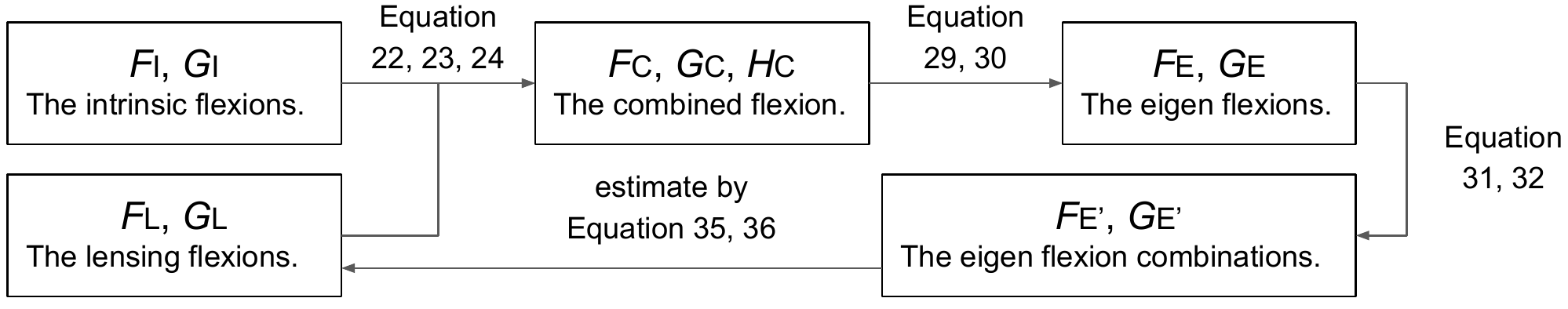}}
\caption{
\label{fig:FlexionC}
The relations between all kinds of the flexions in the steps for measuring the lensing flexions.
}
\end{figure*}

{\bf
At the end of this section, we summarize the steps to measure weak lensing flexion.
The all steps for measuring the lensing flexions are as follows, and figure \ref{fig:FlexionC} shows the relations between all kinds of the flexions.

The first step is to decide the constraint condition which constrains the relation between the combined first flexion and the combined third flexion. 
Although the condition can be chosen from many possibilities, the condition we used in the simulation data, $\mbcF_C=\mbcH_C^*$, seems reasonable because of its simplicity as well as  the result of the simulation test.

The second step is to measure the zero moments of the object image. 
The zero moments are image moments of the object which are defined to have zero value if the spin number of the moments is not zero.
The formalism of the zero moments is defined in the zero plane and it has been transformed in the image plane(equation \ref{eq:zeroM_flexion}).
Thus, the zero moments can be determined in the image plane, 
by measuring five shape parameters of the observed image [centroid, combined shear, three combined flexions] and then by using the constraint equation we can make the four zero moments [$\cM^1_1$, $\cM^2_2$, $\cM^3_1$, $\cM^3_3$]  zero.

The third step is to calculate the eigen flexion combination. 
The combined flexions measured above depend on the constraint condition and include  the intrinsic shear terms, so it is needed to calculate the eigen flexion combinations (equations \ref{eq:eigen_flexion1}, \ref{eq:eigen_flexion2}, \ref{eq:z_flexion1} and \ref{eq:z_flexion2}) to obtain weak lensing flexions precisely.

The final step is to average of the eigen flexion combinations to obtain weak lensing flexions.
The average value is obtained as a combination of lensing shear and lensing flexions (equations \ref{eq:ave_z_flexion1} and \ref{eq:ave_z_flexion2}), so the lensing flexions are obtained by the inverse transformation(equations \ref{eq:WLF1} and \ref{eq:WLF2}) after obtaining lensing shear from average of the combined shear, $\mbg_L=\lrt{\mbg_C}$.
}

\begin{figure}
 \begin{minipage}{0.495\hsize}
  \begin{center}
\vspace{2mm}
   \includegraphics[width=70mm]{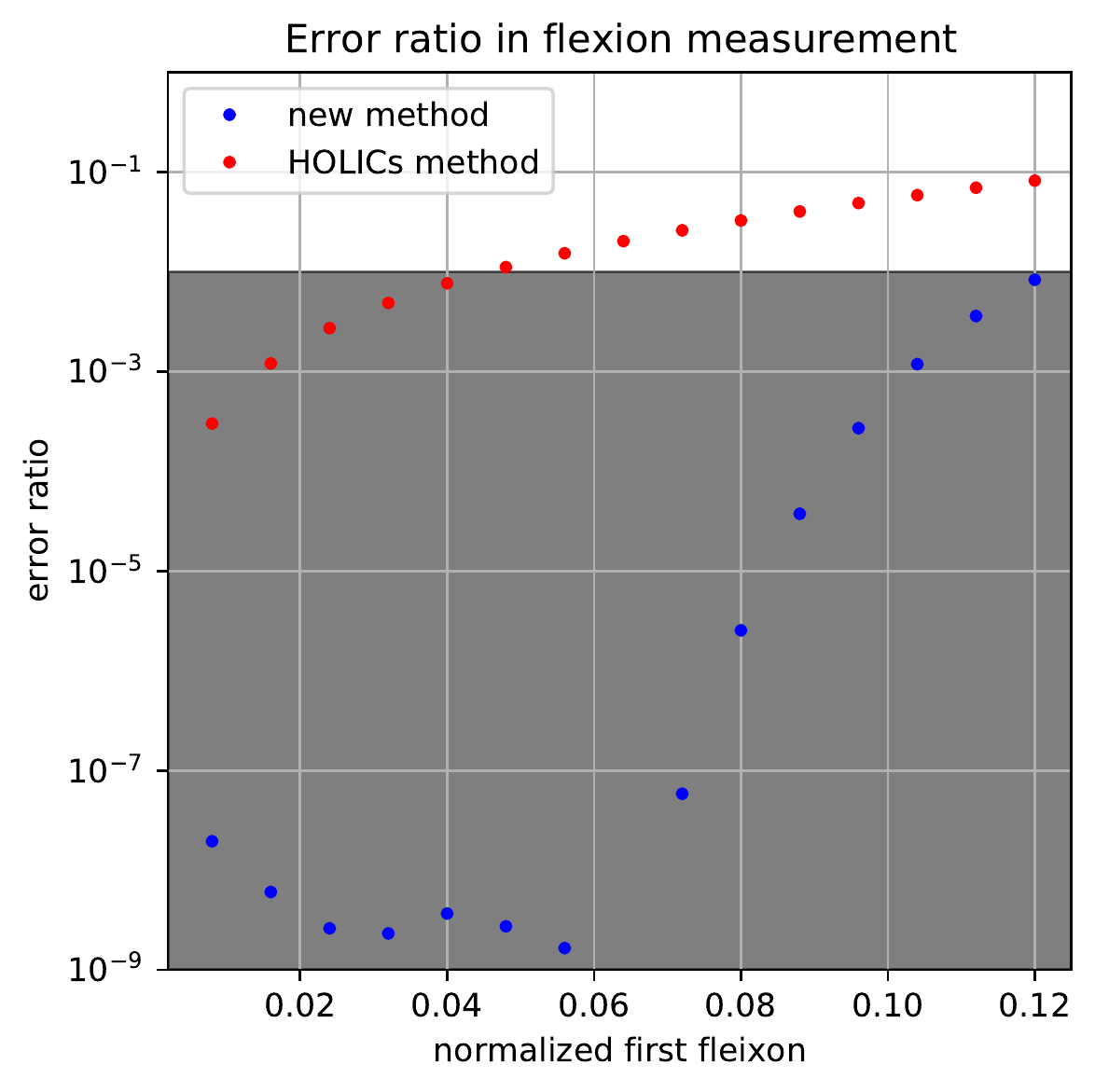}
  \end{center}
  \caption{
The error ratio of the first flexion measurement for distortion set "F only". 
The horizontal axis means the normalized input first flexion which is defined the products of the input first flexion $\mbcF_t$[/pixels] and the Gaussian radius of the simulated image $r_g=4$[pixels] , and the vertical axis means the error ratio $\mbcF_e$, and the gray region means the error ratio is under 1\%.
The blue points mean the new method and the red points mean the HOLICs method.
}
  \label{fig:paper1_2_F_F}
 \end{minipage}
\hspace{3mm}
 \begin{minipage}{0.495\hsize}
  \begin{center}
   \includegraphics[width=70mm]{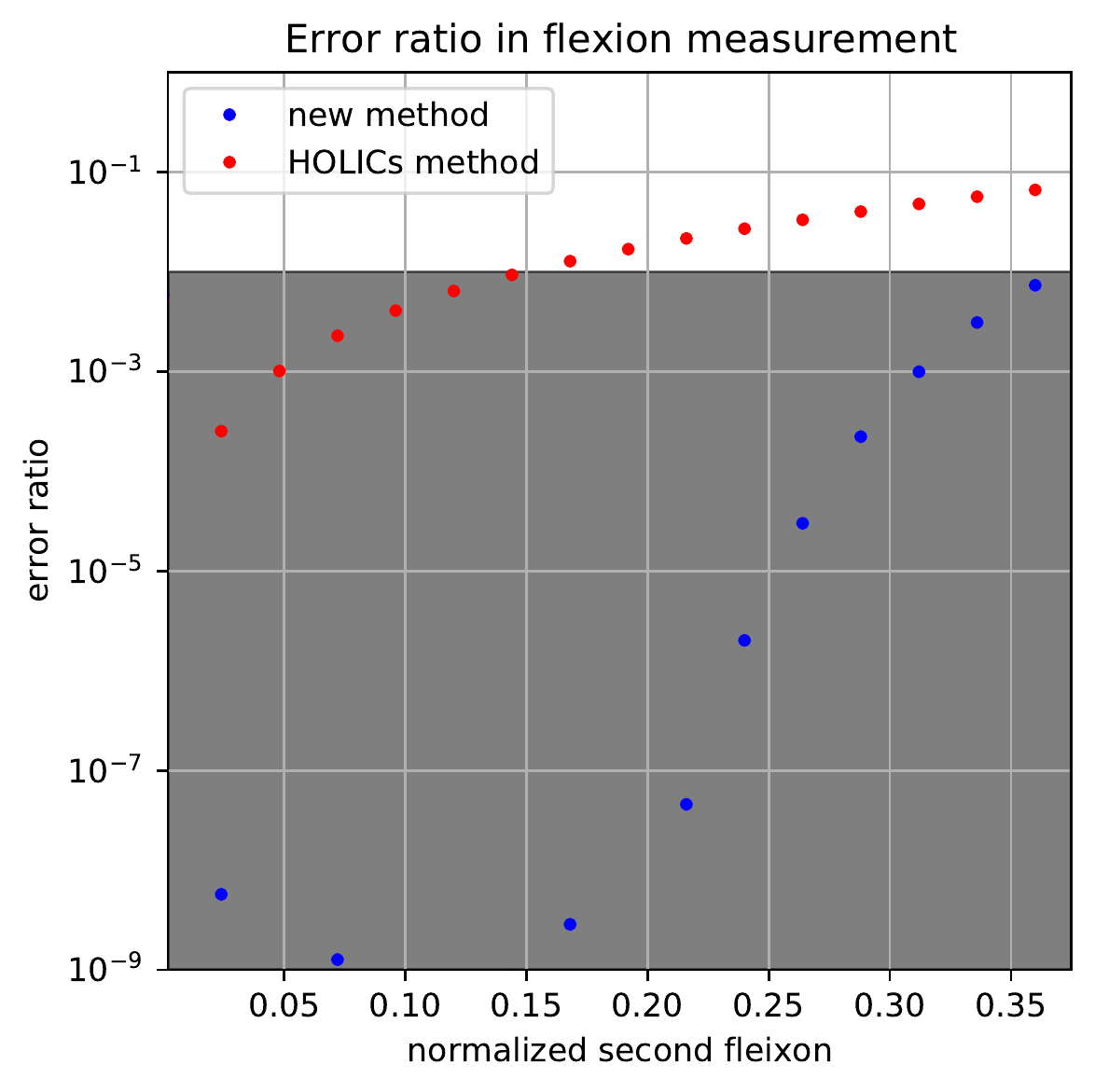}
  \end{center}
  \caption{
Same plot as figure \ref{fig:paper1_2_F_F} except that the vertical axis means the error ratio of the second flexion measurement for distortion set "G only" and the horizontal axis means the normalized input second flexion $\mbcG_t$.}
  \label{fig:paper1_2_G_G}
 \end{minipage}
\end{figure}

\begin{figure}
 \begin{minipage}{0.495\hsize}
  \begin{center}
\vspace{2mm}
   \includegraphics[width=70mm]{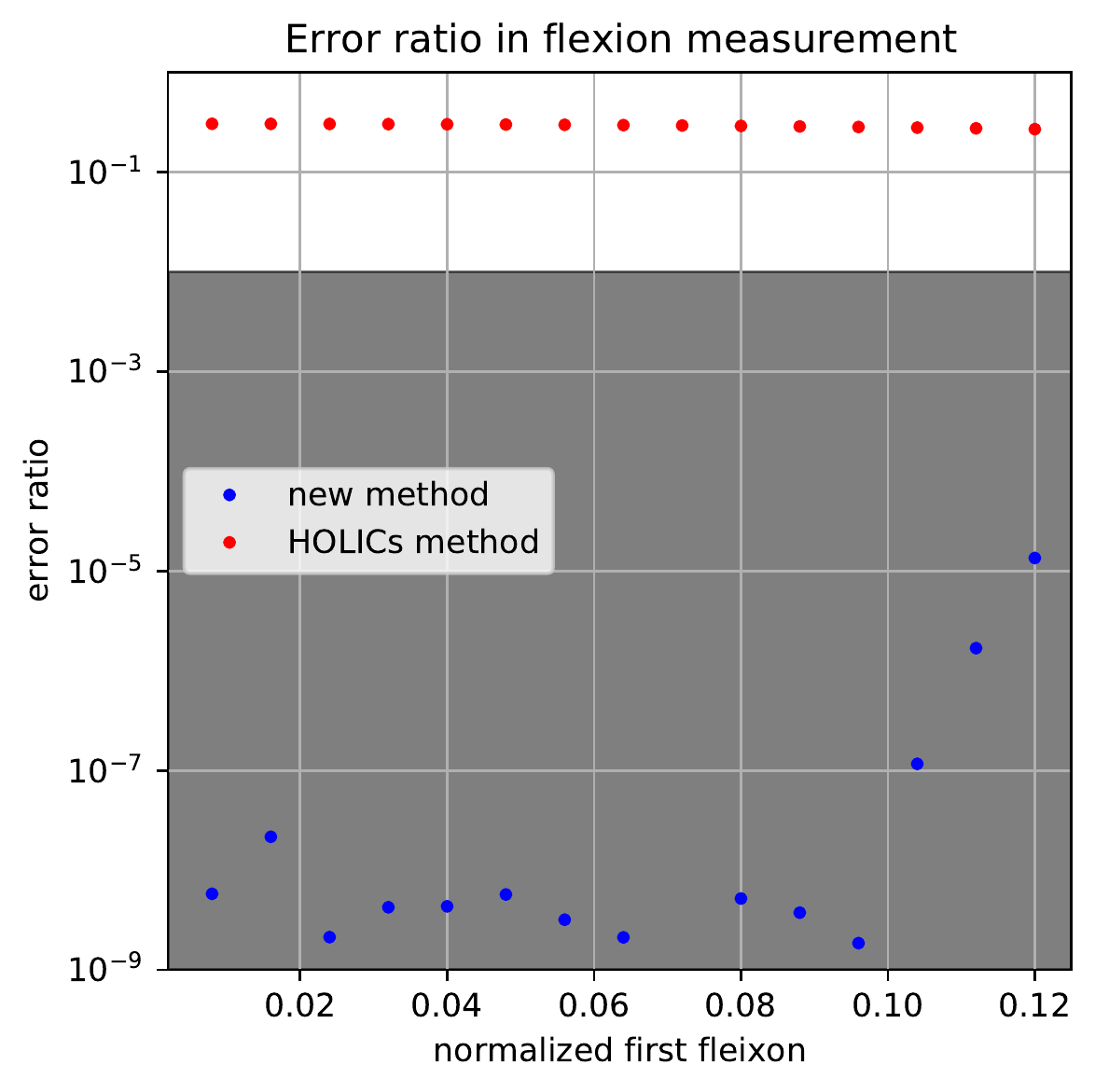}
  \end{center}
  \caption{
Same plot as figure \ref{fig:paper1_2_F_F} except that the vertical axis means the error ratio of the second flexion measurement for distortion set "F and g" and the horizontal axis means the normalized input second flexion $\mbcF_t$.}
  \label{fig:paper1_2_Fg_F}
 \end{minipage}
\hspace{3mm}
 \begin{minipage}{0.495\hsize}
  \begin{center}
   \includegraphics[width=70mm]{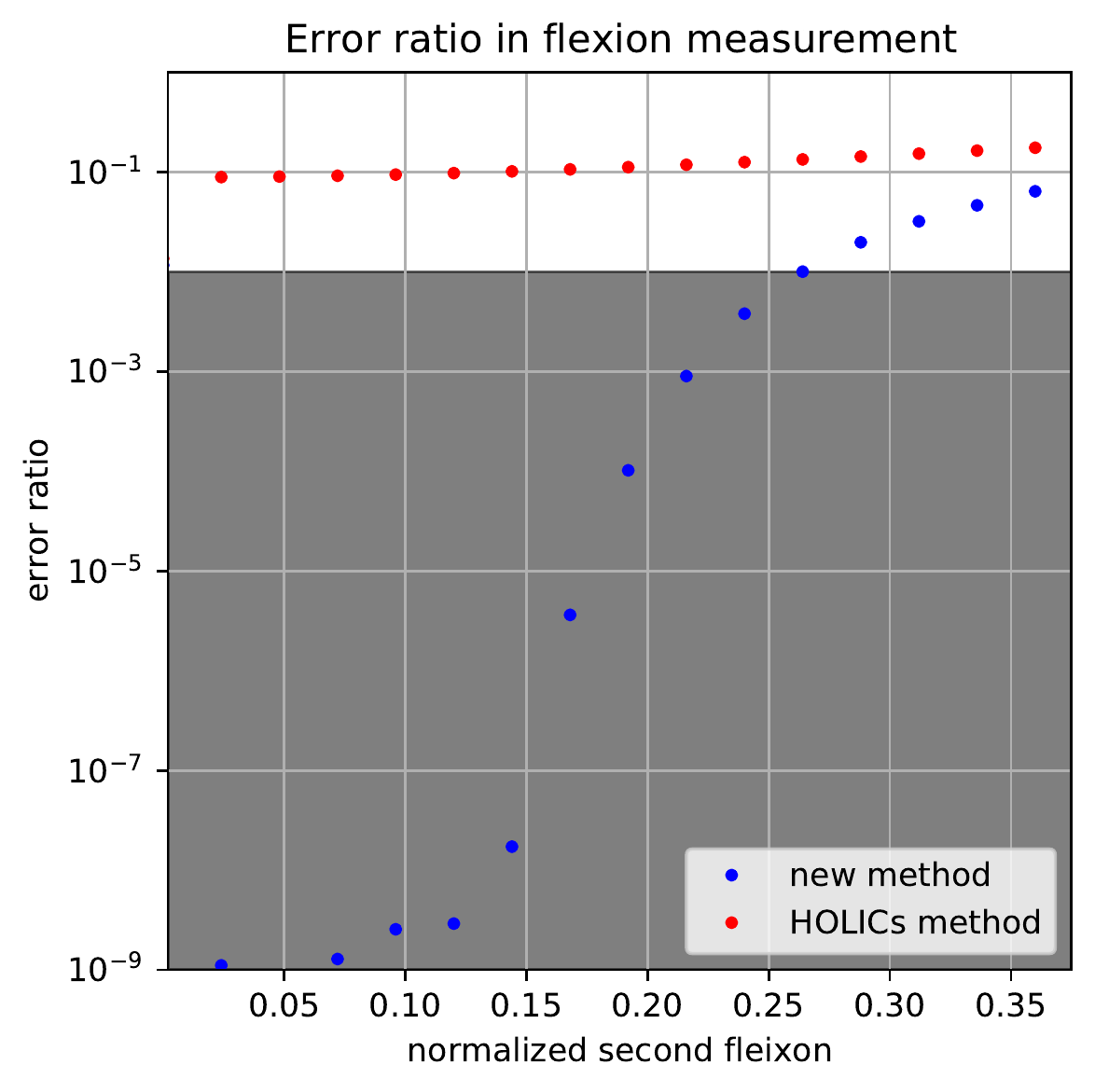}
  \end{center}
  \caption{
Same plot as figure \ref{fig:paper1_2_F_F} except that the vertical axis means the error ratio of the second flexion measurement for distortion set "G and g" and the horizontal axis means the normalized input second flexion $\mbcG_t$.}
  \label{fig:paper1_2_Gg_G}
 \end{minipage}
\end{figure}

{\bf
\subsection{Comparison with HOLICs method}
In this section, we compare the precision of the new flexion measurement method with the HOLICs method which was developed before by us. The details of the HOLICs method can be seen in Okura, Umetsu and Futamase 2007 and 2008.

Because the HOLICs method does not consider the third flexion (equation \ref{eq:lenseq_flexion3}), to investigate the difference of pure shape measurement error, we compare the two methods with a simple situation that has only lensing distortion.
We made four simulation images with Gaussian profile and Gaussian radius = 4 pixels and distorted them with a different distortion set.
The each distortion set have distortion parameters as 
the first flexion only($\mbg_L=0.0$, $\mbcG_L=0.0$: "F only"), 
the second flexion only($\mbg_L=0.0$, $\mbcF_L=0.0$: "G only"), 
the first flexion with ellipticity($\mbg_L=-0.3$, $\mbcG_L=0.0$: "F and g") and 
the second flexion with ellipticity($\mbg_L=-0.3$, $\mbcF_L=0.0$: "G and g").
The precisions of each method are compared with the error ratio $\mbcF_e$ or $\mbcG_e$ of flexion measurement, 
the error ratio is defined from the true value $\mbcF_L$ or $\mbcG_L$ used for making the simulation image and measured value $\mbcF_m$ or $\mbcG_m$ as
\begin{eqnarray}
\label{eq:error_ratio}
\mbcF_e &=& \lrv{\frac{\mbcF_m-\mbcF_L}{\mbcF_L}}
\\
\mbcG_e &=& \lrv{\frac{\mbcG_m-\mbcG_L}{\mbcG_L}}.
\end{eqnarray}

The figure \ref{fig:paper1_2_F_F} and \ref{fig:paper1_2_G_G} show the result of the comparison with the distortion set "F only" and "G only".
We can see the error ratio of the new method is very small even in a strong flexion region which makes the error ratio more than 1\% in the HOLICs method.
The figure \ref{fig:paper1_2_Fg_F} and \ref{fig:paper1_2_Gg_G} show the result of the comparison with the distortion set "F and g" and "G and g".
We can see the HOLICs method has a large error ratio even in weak flexion limits, because the HOLICs method does not consider the combination of shear and flexion.
However, the new method has a very small error ratio even when shear and flexion are combined, because the new method considers the effect from shear in flexion measurement.
These results show clearly that the new method is an improved method much from the HOLICs method.
}

\section{Summary}
\label{sec:Summary}
In this paper, we developed  a new precise flexions measurement method.
Flexions appeared in third order terms in the expansion of the lensing equation than shear,
so the formulas to measure flexions are much more complicated than that of shear.

We applied the idea of the zero plane to calculate the relation between the image plane and the zero plane.
In the course of derivation, we have  a third type of flexion which is independent from the first and the second flexions.
This causes a very serious problem, because the addition of independent parameters makes it impossible to measure the three flexions uniquely.
So we have to use some constraint condition to determine uniquely flexion,
but it means the measured flexion and average of the flexion depend on the selection of the constraint condition.
We found that the distorted image from circular shape has only two combinations of combined flexions, called eigen flexions.
It is found that the eigen flexion is independent from the constraint condition.
But the average of the eigen flexion includes intrinsic noise which cannot be eliminated by averaging.
We then constructed the combinations of eigen flexions called eigen flexion combinations which contain only linear terms of the intrinsic distortion.
After averaging the eigen flexion combination the dependence on the intrinsic value can be eliminated.
The average of the new combinations still contain the lensing shear but it can be obtained from the average of
the combined shear.
We confirmed using a simple numerical simulation that the average of the eigen flexion combination do not depend on
the choice of the constraint and are free from the intrinsic noise.

In this way we have developed a method which can measure the lensing flexion by measuring shape of galaxy image,
but there are other effects in real observation  which change the shape of the image on the top of the lensed image.
We need to correct them to have the lensed image before applying the method of flexion analysis developed above.
One of the most important corrections is PFS correction which we will study in the next paper.


\appendix
\section{Additional mask for moment measurement} 
\label{sec:Mask}
In this section we explain about the additional mask for weight function in moment measurement.

\begin{figure*}
\centering
\resizebox{0.95\hsize}{!}{\includegraphics{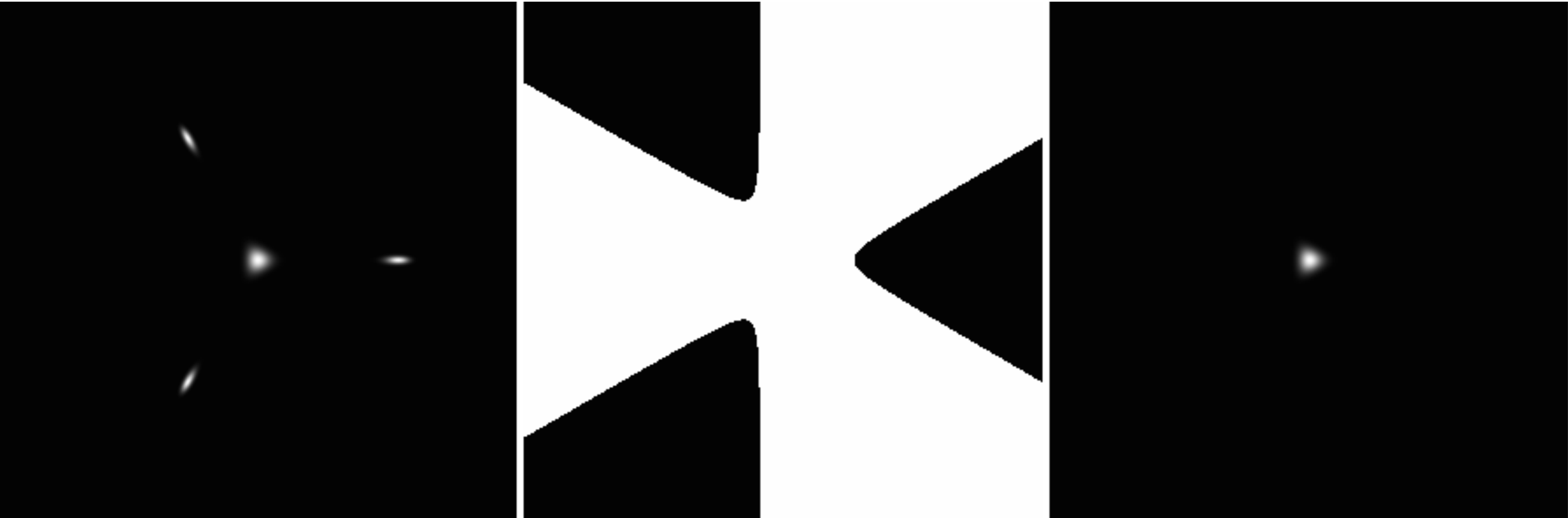}}
\caption{
\label{fig:MaskG}
Weight(left), mask(center) and masked weight(right) region for image with 2nd flexion only.
The black region in the center image is a masked region.}
\end{figure*}

\begin{figure*}
\centering
\resizebox{0.95\hsize}{!}{\includegraphics{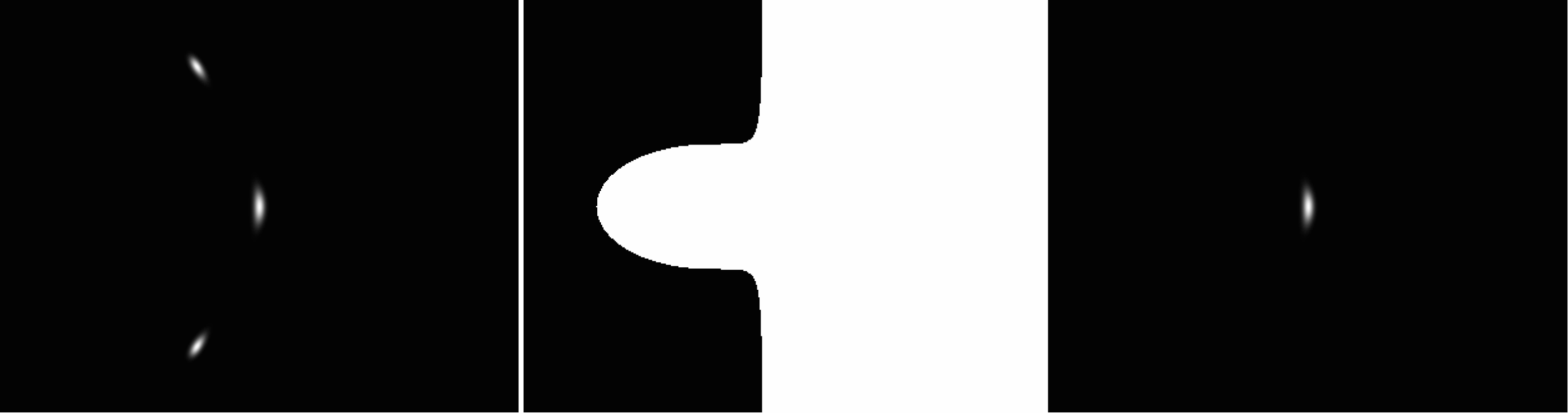}}
\caption{
\label{fig:MaskF}
Same image as figure \ref{fig:MaskG} except the distortion is shear and both flexions.}
\end{figure*}

As we explained in section \ref{sec:Zero moments and shapes}, a weight function which is defined as a circular function in the zero plane is used for the moment measurement.
The variable of the weight function is simply $|d\tbbe|$ in the zero plane, but it has nonlinear terms in the image plane and the terms make the distribution of the weight function a very complicated.
The left image of figure \ref{fig:MaskG} and figure \ref{fig:MaskF} are Gaussian weight functions in the image plane with flexion distortions.
We can see there are some peaks in the outer part from the central peak where the object is located.
If there are other objects on the peaks, the weak lensing information can not be obtained correctly.
{\bf
The peaks appear even in weak flexion limits.
Because the position of the peaks has a tendency to be far from the object in weak flexion and close to the object in stronger flexion, so in some situations, e.g. weak flexion or taking a small stamp for the object, it may be able to be ignored, however it is clear that the peaks can be ignored in all cases of future shape measurement.
One of the reliable techniques to remove the peak is using a mask for the weight function.
} 
Because the peaks are in the outer region with ${\delta\tbbe(\bth)\cdot d\bth}>0$, the appropriate mask $M(\bth)$ can be obtained as
\begin{eqnarray}
\label{eq:Mask}
M(\bth) &\equiv& H\lr{\delta\tbbe(\bth)\cdot d\bth}\\
\delta\tbbe(\bth)& \equiv& \PP{\lr{d\tbbe}}{\lr{d\bth}}d\bth +  \PP{\lr{d\tbbe}}{\lr{d\bth^*}}d\bth^*,
\end{eqnarray}
where $H$ is Heaviside step function.
The center images of the figures are the mask regions calculated using equation \ref{eq:Mask},
then the right images of the figures are masked weight.
We can see the Masked weight can mask the peaks.


\section*{Data availability}
The simulation data underlying this article were created my original python codes, so these are available on request.

\section*{Acknowledgements}
This work is supported in part by a Grant-in-Aid for Science Research from
JSPS (No.20K03937 to T.F).


\end{document}